\documentclass{article}

\usepackage[english]{babel}

\usepackage[a4paper,top=2cm,bottom=2cm,left=3cm,right=3cm,marginparwidth=1.75cm]{geometry}
\usepackage{changepage}
\usepackage[colorlinks=true, allcolors=blue]{hyperref}
\newcommand{\slightlylarge}{\fontsize{11pt}{13pt}\selectfont}

\newcommand{\glossentry}[4]{%
  \vspace{0.8em}
  \noindent\textbf{\slightlylarge #1}: #2\par
  \vspace{0.4em}
  \begin{adjustwidth}{2em}{0pt}
    {\hangindent=10em\hangafter=1 \noindent\textbf{Example:} #3\par\vspace{0.4em}}
    {\hangindent=10em\hangafter=1 \noindent\textbf{Why it matters:} #4\par}
  \end{adjustwidth}
}

\usepackage{booktabs} 
\usepackage{amsmath}
\usepackage{graphicx}
\usepackage[colorlinks=true, allcolors=blue]{hyperref}
\usepackage{authblk}
\usepackage[svgnames]{xcolor} 
\usepackage{tcolorbox}
\tcbuselibrary{skins} 

\tcbset{
  pillstyle/.style={
    enhanced,         
    arc=1.5mm,        
    boxrule=0pt,      
    coltext=white,    
    left=1pt,         
    right=1pt,        
    top=1.5pt,        
    bottom=1.5pt,     
    fontupper=\sffamily\bfseries\scriptsize, 
    nobeforeafter,    
    tcbox raise base, 
  }
}

\newcommand{\DE}{\tcbox[pillstyle, colback=SteelBlue]{DE}}
\newcommand{\DS}{\tcbox[pillstyle, colback=DarkSeaGreen]{DS}}
\newcommand{\MLE}{\tcbox[pillstyle, colback=LightSeaGreen]{MLE}}
\newcommand{\ADE}{\tcbox[pillstyle, colback=MediumOrchid]{AE}}
\newcommand{\BA}{\tcbox[pillstyle, colback=Tan]{BA}}
\newcommand{\GO}{\tcbox[pillstyle, colback=SlateGray]{GO}}
\newcommand{\DME}{\tcbox[pillstyle, colback=Sienna]{DME}}
\newcommand{\PM}{\tcbox[pillstyle, colback=Peru]{PM}}
\newcommand{\LG}{\tcbox[pillstyle, colback=LightPink]{LG}} 

\definecolor{valueconnectioncolframe}{rgb}{0.9, 0.9, 9}
\definecolor{valueconnectionbgcol}{rgb}{0.95, 0.95, 0.95}
\definecolor{dataprocesscol}{HTML}{FFC000}
\definecolor{modelprocesscol}{HTML}{46B1E1}
\definecolor{inferenceprocesscol}{HTML}{F2AA84}
\definecolor{applicationprocesscol}{HTML}{00CC99}
\definecolor{monitoring}{HTML}{000000}

\newtcolorbox{valueconnection}[2][]{
    colback=#2,
    colframe=#2,
    spread outwards,
    colbacktitle=#2,
    coltitle=white,
    coltext=white,
    titlerule style=#2,
    fonttitle=\large\bfseries,
    title=#1 value connection,
    enhanced,
    arc=0mm,
    boxrule=0.25mm,
    toptitle=5mm,
    bottom=0.5cm,
    left=1cm,
    right=2cm
}

\title{AI Application Operations - A Socio-Technical Framework for Data-driven Organizations}
\author[1]{Daniel Jönsson}
\author[1]{Mattias Tiger}
\author[2]{Stefan Ekberg}
\author[3]{Daniel Jakobsson}
\author[4]{Mattias Jonhede}
\author[1]{Fredrik Viksten}
\affil[1]{Linköping University}
\affil[2]{AI Sweden}
\affil[3]{Swedish Transport Administration}
\affil[4]{Volvo Group}

\begin{document}
\maketitle

\begin{abstract}
We outline a comprehensive framework for artificial intelligence (AI) Application Operations (AIAppOps), based on real-world experiences from diverse organizations. 
Data-driven projects pose additional challenges to organizations due to their dependency on data across the development and operations cycles. 
To aid organizations in dealing with these challenges, we present a framework outlining the main steps and roles involved in going from idea to production for data-driven solutions. 
The data dependency of these projects entails additional requirements on continuous monitoring and feedback, as deviations can emerge in any process step.
Therefore, the framework embeds monitoring not merely as a safeguard, but as a unifying feedback mechanism that drives continuous improvement, compliance, and sustained value realization—anchored in both statistical and formal assurance methods that extend runtime verification concepts from safety-critical AI to organizational operations.
The proposed framework is structured across core technical processes and supporting services to guide both new initiatives and maturing AI programs.
\end{abstract}

\section{Introduction}
AI systems are increasingly being deployed in critical domains where transparency, safety, and performance must be maintained across the full model lifecycle. To support sustainable and trustworthy AI adoption, organizations need clearly structured processes that span data, model development, deployment, and operations. This paper presents a framework for AI Application Operations (AIAppOps), identifying common challenges and effective practices at each stage, with a particular emphasis on how continuous monitoring bridges the technical and organizational dimensions of AI practice.
The framework has been developed based on expert interviews, workshops, and surveys within a large consortium of companies\footnote{Aixia, IBM, NetApp, Predli, Proact IT Sweden, Stormgrid, Volvo Group, Red Hat, Hopsworks}, government agencies \footnote{Swedish Traffic Agency, Sahlgrenska University Hospital, Swedish Tax Agency, Statistics Sweden (SCB), Region Halland and Västra Götaland regional health care providers}, and research institutes\footnote{Linköping University, Research Institutes of Sweden (RISE), Santa Anna IT Research Institute} transforming into, or working with, data-driven organizations.  
Thus, it integrates perspectives from industry, government, and academia to facilitate alignment across organizational boundaries.
\\\\
\noindent
\textbf{Scope and contribution.}
This white paper consolidates the collective experience of a multi-stakeholder consortium spanning public agencies, private companies, and research institutions. It is written as a \emph{practitioner–research synthesis}: bridging empirical observations from real AI deployments with concepts from academic and industrial literature on machine learning operations (MLOps), data readiness, and AI governance.
Whereas most prior works focus either on technical MLOps pipelines or isolated data-preparation phases, this paper proposes a coherent, end-to-end framework that spans the full lifecycle from idea to sustained value in production.
A defining difference between AI-driven development and traditional software engineering is the inherently exploratory nature of machine learning. In conventional software projects, organizations can usually predict the functional outcome of an investment with high confidence once requirements are specified. In contrast, AI projects require organizations to formulate and validate bets: data is explored, models are built, and only through empirical validation can teams determine whether the approach is viable and whether the expected value can be realized. This exploratory character fundamentally changes how organizations must reason about risk, progress, and success.
In this respect, the work extends the canonical MLOps perspective by explicitly including (i) the \textbf{application and usage} stages—how models are integrated into real systems and continuously assessed in use—and (ii) the \textbf{value-driven meta-process} that links technical observability to business, ethical, and regulatory outcomes. 

The white paper therefore aims to serve three audiences simultaneously:
\begin{itemize}
\item \emph{Practitioners}, who can adopt the described processes as templates for reliable and compliant AI delivery;
\item \emph{Researchers}, who can view the framework as an empirical extension of existing MLOps and data-readiness theories; and
\item \emph{Policy makers and organizational leaders}, who can recognize how operational practices implement the principles of trustworthy and value-aligned AI.
\end{itemize}

\begin{figure}[!ht]
\centering
\includegraphics[width=\linewidth]{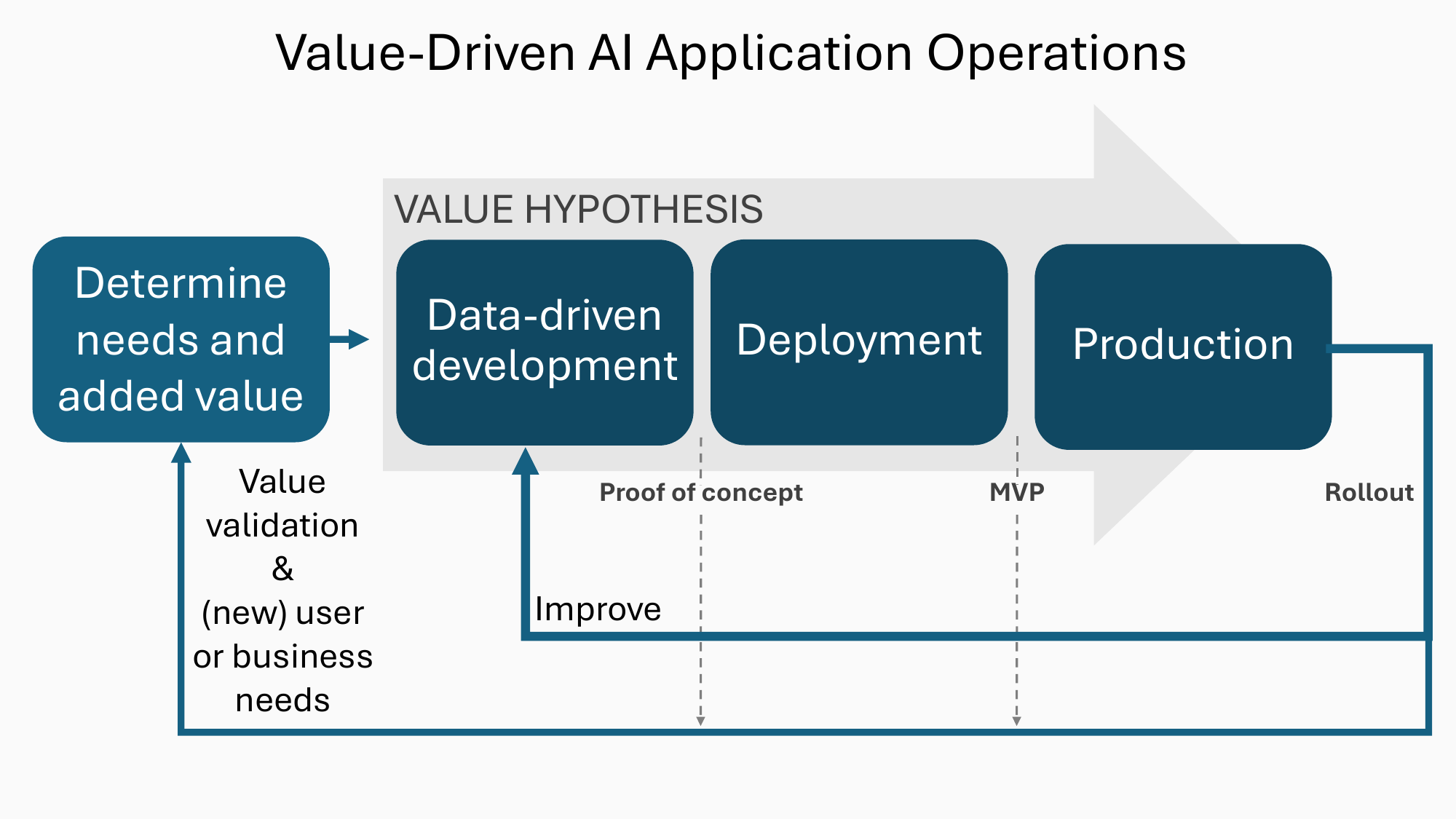}
\caption{\label{fig:Value-driven-AI-application-operations}Illustration of the process for value-driven AI application operations, starting with determining business needs and finally using production-generated data for developing new applications or improving the existing ones. The process is inspired by the phases of CRISP-DM\cite{chapman2000crispdm}, but extends them to also take the feedback from deployed operations into account.}
\end{figure}

The result is an explicitly \emph{value-driven framework}; every technical activity from data curation to application integration is anchored in a declared \emph{value hypothesis}\footnote{For clarity, all key terms,  abbreviations and concepts used throughout this paper are defined in the \hyperref[sec:glossary]{Glossary} at the end.} and advanced only when evidence supports that hypothesis, see \autoref{fig:Value-driven-AI-application-operations}. Concretely, teams start by stating the intended outcomes (e.g., cost reduction, error avoidance, service-level improvements, user satisfaction, learning outcomes), the proxies that can be measured early, and the guardrails (safety, legal, ethical) that must hold for the value to be deployable. Delivery then proceeds in stages (proof-of-concept $\rightarrow$ MVP $\rightarrow$ rollout), each with pre-defined thresholds that gate the next step.
Importantly, these thresholds are provisional: true value can only be validated once the application is embedded in real operational contexts. The framework therefore treats value validation as a recurring activity rather than a one-time decision, with late-stage and post-deployment feedback carrying the highest evidentiary weight.
Operations maintain and increase value through continuous monitoring, analysis, and change, feeding downstream signals back into upstream work. This feedback-driven operating mode represents a cultural shift for many organizations: success is no longer defined by initial delivery, but by sustained learning and adaptation under uncertainty.

This stance both contextualizes and \emph{constrains} process choices. For example, if downstream value depends on safe human triage, the model must provide calibrated uncertainty or abstain (“I don't know”) under distributional shift. If value must be \emph{defensible} in regulated contexts, explainability, lineage, and auditability become non-negotiable non-functional requirements. Likewise, instrumentation for value (business KPIs, task metrics, and safety indicators) is designed \emph{before} building, so that experiments can be judged against the hypothesis. 

The framework applies to both predictive and generative AI. Predictive systems typically articulate value in terms of decision quality and workload shaping (e.g., avoided false negatives, time-to-resolution), with calibration and cost-sensitive metrics tied to business outcomes. Generative systems articulate value in terms of task completion and quality (e.g., deflection rate, handling time, citation fidelity, refusal behavior), with policy and safety guardrails integrated into evaluation and serving. In both cases, value realization depends on the \emph{application}—how predictions or generations change real workflows—not solely on model accuracy nor precision.
\section{Related Work}

Machine Learning Operations (MLOps) has emerged as the principal framework for managing the lifecycle of machine learning models.
Google’s \emph{Practitioners Guide to MLOps}~\cite{google2021mlops} and subsequent definitions such as Kreuzberger et al.~\cite{kreuzberger2023mlops} describe standardized processes and capabilities—experiment tracking, model training, deployment, and monitoring—aimed at reliable automation and reproducibility.
These frameworks typically emphasize \textbf{automation, scalability, and reproducibility} but focus primarily on the model and infrastructure layers.
Our work builds directly on this foundation but extends the scope beyond model deployment to include the surrounding \emph{application}, \emph{value realization}, and \emph{governance} layers of AI systems.

Several complementary strands of literature address adjacent gaps.
\textbf{Data readiness and data-centric AI} have gained renewed attention with works such as Lawrence’s \emph{Data Readiness Levels}~\cite{lawrence2017drl}, the \emph{Data Readiness Report}~\cite{afzal2021data}, and the recent extensions by Tiger et al.~\cite{tiger2024eva} and Jakubik et al.~\cite{jakubik2024dca}.
These emphasize systematic data preparation and quality documentation as prerequisites for reliable modeling.
AIAppOps incorporates these insights by structuring the \emph{Data Process} into model-agnostic curation and model-specific adaptation stages, thereby operationalizing data readiness within the continuous lifecycle.

A growing body of work examines \textbf{monitoring and maintenance} of deployed models.
Naveed et al.’s review~\cite{naveed2025monitoring} shows that most monitoring practices still focus narrowly on technical metrics (e.g., latency, accuracy, drift), while aspects such as fairness, privacy, and business impact, the focus of this work, remain underrepresented.
At the theoretical end, Tiger’s work on safety-aware autonomous systems~\cite{tiger2020probstl,tiger2023safetyaware} formalizes monitoring as runtime verification under uncertainty, introducing Probabilistic Signal Temporal Logic (ProbSTL)\footnote{ProbSTL (Probabilistic Signal Temporal Logic) formalizes uncertainty-aware runtime monitoring by expressing probabilistic temporal constraints over continuous signals. Developed by Tiger~\cite{tiger2020probstl,tiger2023safetyaware}, it provides the mathematical underpinning for AIAppOps’ highest maturity level of monitoring.}
as a method for reasoning over probabilistic events and temporal guarantees.
These contributions extend monitoring beyond metric collection to formal assurance, a concept that AIAppOps generalizes into socio-technical observability across the full AI lifecycle.
The AIAppOps framework explicitly addresses these gaps by linking technical observability to value and trustworthiness indicators, ensuring that \emph{data, model, inference, and application} are monitored jointly.

Finally, recent policy developments, particularly the \emph{EU Artificial Intelligence Act}~\cite{euai2024}, stress continuous post-deployment monitoring, documentation, and human oversight for high-risk AI systems.
The governance and compliance mechanisms embedded throughout AIAppOps provide a practical way to meet these regulatory expectations.
\\\\
\noindent In summary, AIAppOps extends MLOps and related data-centric frameworks by introducing:
\begin{enumerate}
\item a value-driven meta-process aligning technical and organizational outcomes,
\item explicit processes for application integration and real-world usage, and
\item cross-cutting governance ensuring lawful, ethical, and sustainable operation.
\end{enumerate}

\section{Method}
The core of our framework follows previous works on machine learning operations (MLOps)~\cite{chapman2000crispdm, kreuzberger2023mlops}, but takes a wider perspective by integrating value-driven and governance-oriented processes that extend beyond the traditional model lifecycle.

\subsection{AIAppOps as an Extension of MLOps}
MLOps provides the foundational practices for automating the development, deployment, and monitoring of machine learning models~\cite{kreuzberger2023mlops,google2021mlops}.
However, most MLOps frameworks stop at the point where a model is deployed and technically monitored, e.g., \cite{spjuth2021machine, kreuzberger2023mlops}.
AI Application Operations (AIAppOps) extends this lifecycle in two essential dimensions:

\begin{itemize}
\item \textbf{Downstream extension:} adding the \emph{Application} and \emph{Usage} phases, which address how models are integrated into software systems and operational decision flows, and how AI capabilities are implemented within business processes so that value can be measured and sustained in real-world operation.
\item \textbf{Cross-cutting extension:} embedding \emph{value alignment, trustworthiness, and compliance} as continuous feedback loops across all phases—from data to application—rather than as external audits.
\end{itemize}
Here, \emph{usage} should be understood broadly to include both human-facing interaction and fully automated decision execution. Crucially, successful value realization depends not only on technical integration, but on the maturity of the surrounding business process: roles, trust, incentives, and decision authority must evolve for AI-supported or AI-automated decisions to have effect.

In contrast to purely model-centric approaches, AIAppOps defines operational responsibilities for application developers, business owners, domain experts, and legal officers alongside data scientists and ML engineers.
This broadens the scope of MLOps into a \emph{socio-technical framework} that ensures not only that models are deployed correctly, but that AI applications continue to deliver intended outcomes safely, lawfully, and sustainably.

\subsection{Workshops}
We performed a series of workshops with organizations at various AI maturity levels, please refer to \autoref{sec:organizational_maturity} for maturity level assessment. 
Each workshop started with a description of the organization's current process with concrete examples of past and ongoing data-driven projects. 
The process description session was followed by discussions about the challenges they faced in their current stage and which challenges they saw in the near future.
We then adapted and aligned our process description to capture best practices from both theirs and previously interviewed organization's data-driven processes.
This allowed us to come up with a process description that captured both best practices described in previous literature~\cite{kreuzberger2023mlops, shankar2024we} and the current workflow of the interviewed organizations.

After the initial workshops with the organizations, we discussed the resulting process description to ensure that the abstraction levels were appropriate and clear. Finally, we presented the process description to a larger number of new organizations to ensure that it captured and covered the needs in diverse sets of AI applications and organizations.
Comments and discussions after the presentation were integrated into the process description, resulting in the AI Application Operations process presented in the following section.

\section{AI Application Operations}
\begin{figure}[ht]
\centering
\includegraphics[width=\linewidth]{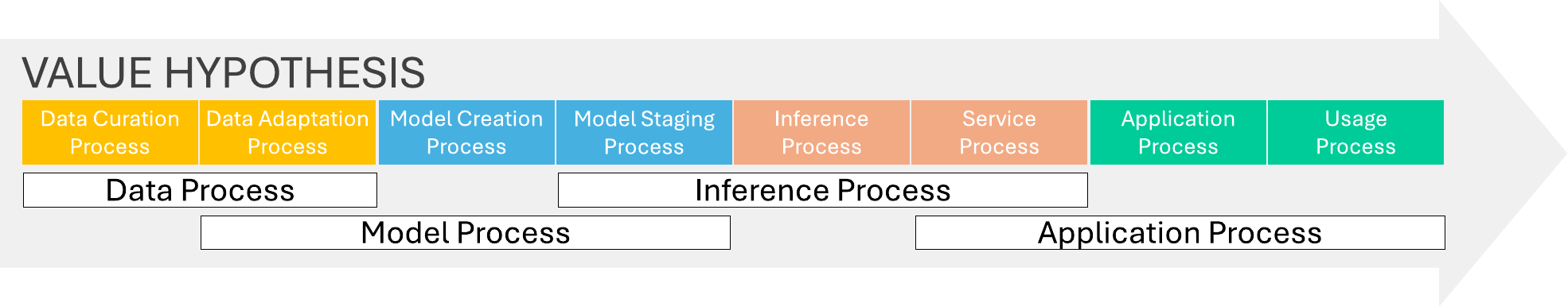}
\caption{\label{fig:process-overview-details}A layered view of the processes in the AI application life-cycle with the value hypothesis as driving force and means for validation. In addition to the traditional processes, data, model, and inference (serving), we also highlight the application and its usage due to their importance for the value hypothesis. Typically, all processes and process levels are iterative and continuous due to data updates and knowledge about usage.
}
\end{figure}

We divide the AI application life-cycle into four major processes illustrated in \autoref{fig:process-overview-details}: Data, Model, Inference, and Application. 
Here, the data process makes data ready for model development, the model process transforms curated data into models, 
the inference process deals with reliably serving inferences based on requests to models, and the application process deals with the more traditional development operations using data-driven services.
The major processes are often not completely separated, which we indicate through the overlap between the processes with the in-between steps: data adaptation, model staging, and service provision.
Furthermore, the overlap also highlights the need for inter-team communication and overlapping roles between the steps of the AI application operations. 
 
Here, data adaptation involves model-dependent transformations, model staging validates models for real-world usage, and service provisioning orchestrates chains of inferences as a service to the application.

The linear depiction of the process in \autoref{fig:process-overview-details} reflects that all steps are required for a deployed solution. The linear process should not be interpreted as a one-directional work-flow as it is common with iterations between the sub processes. 
For example, to acquire and adapt new data based on identified needs of the application.  
The value hypothesis (c.f. \autoref{sec:value-driven-operations}) highlights the link between the technical side and the business that needs to be integrated across all processes. 
Support functions such as monitoring, improvement cycles, and compliance are crucial to provide this link and maintain application integrity over time. Subsection \ref{sec:monitoring} details how monitoring operationalizes this continuity by integrating data, model, and application observability into a single feedback loop.

\subsection{Organizational Maturity and Its Impact on AI Operational Processes}\label{sec:organizational_maturity}
Few organizations begin their AI journey with a fully developed process that encompasses the entire AI application life-cycle. Instead, most start with isolated projects that serve as learning experiences and stepping stones. These initial efforts help build the necessary competencies and process components required to eventually manage AI applications systematically and at scale.

\begin{figure}
\centering
\includegraphics[width=\linewidth]{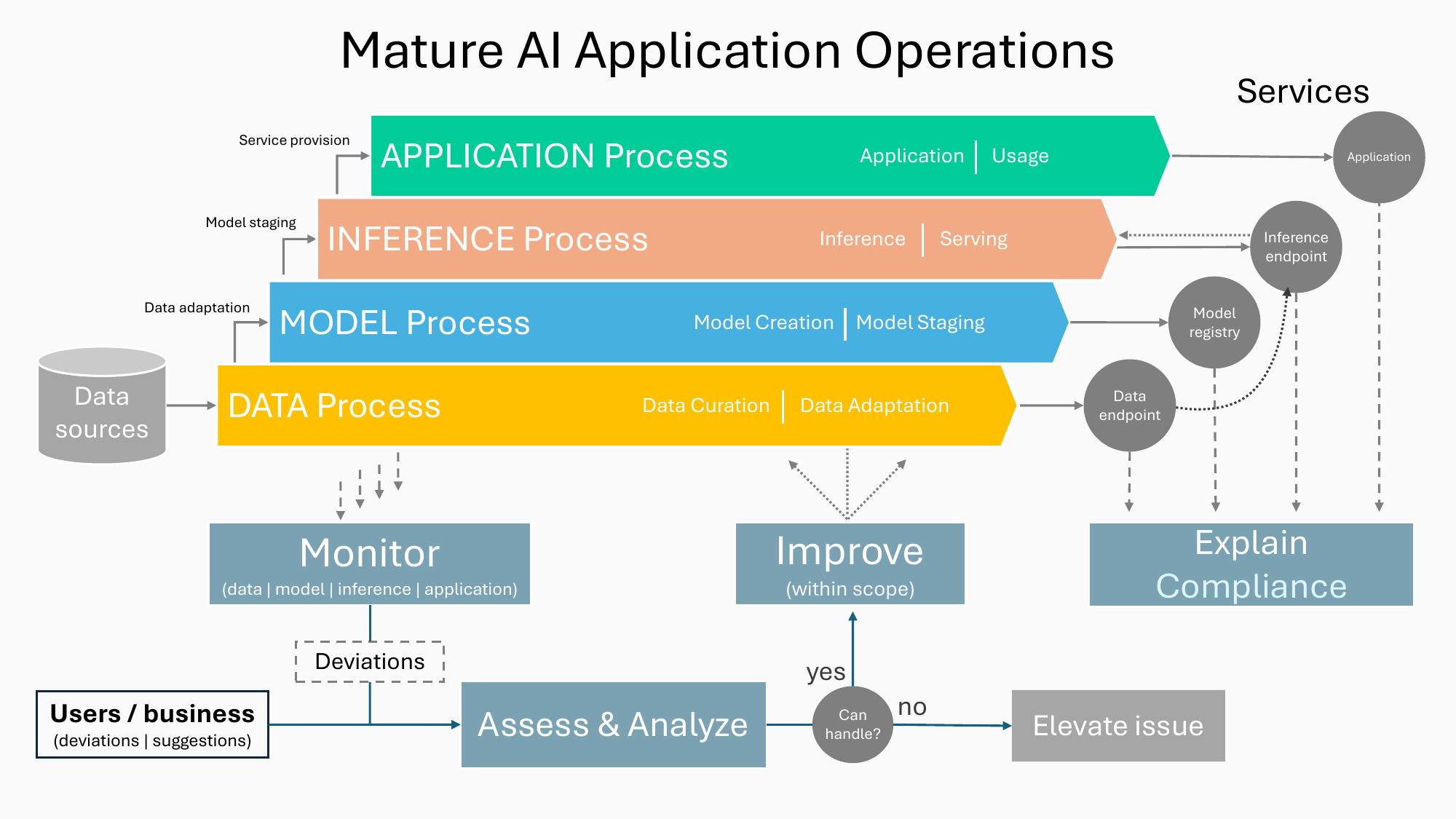}
\caption{\label{fig:AI-application-operations-process-workflow}High-level map of  AI application operations for an organization with high machine learning operations maturity. 
The four core operational processes—data, model, inference, and application—each produce composable endpoints used in downstream services. Monitoring and feedback mechanisms support continuous improvement, while compliance and explainability services ensure operational trustworthiness. The framework supports both proactive and reactive governance by linking technical observability to business-facing outcomes (the value hypothesis, \autoref{sec:value-driven-operations}), and by embedding continuous monitoring as the connective tissue between all lifecycle processes.}
\end{figure}

Once one or a few AI applications are successfully deployed in production, it becomes meaningful to strive for a more complete and integrated operational process. As a reference point, we provide a distilled process map (see \autoref{fig:AI-application-operations-process-workflow}) that illustrates how a mature organization may manage AI application operations. This process includes not only the delivery of AI-powered services but also the organizational structures and feedback mechanisms that enable continuous learning, monitoring, and reuse of data and models across applications.

It is important to emphasize that implementing the entire AI application operations process from the outset is rarely feasible nor advisable. The ability to manage the full AI life-cycle—including continuous deployment, monitoring, and feedback loops—depends heavily on the organization's operational maturity. 

\subsubsection{Assessing Organizational Maturity}
Before starting to transform the organization to AI application operations one should assess the current maturity of the organization~\cite{john2025empirical}.
There are several scales that can guide organizations in evaluating where they stand in their journey from ad-hoc experimentation to enterprise-wide AI operations, e.g., \cite{john2021towards, microsoftMLOPSMaturity}. Here, we use the 5-level MLOps maturity scale defined by Microsoft~\cite{microsoftMLOPSMaturity}. These can be briefly summarized as:

\begin{itemize}
    \item \textbf{Level 0 - No MLOps:} Manual, script-driven processes with no tracking of model performance. Difficult to manage the full ML lifecycle, and releases are painful.
    
    \item \textbf{Level 1 - DevOps, No MLOps:} Application code is managed with automated builds and tests, but model training remains manual. Feedback from production is limited.
    
    \item \textbf{Level 2 - Automated Training:} Fully managed training environment with centralized model tracking and reproducible results. Manual but low-friction releases.
    
    \item \textbf{Level 3 - Automated Model Deployment:} Automatic, low-friction releases with full traceability and integrated A/B testing. Entire environment managed from training to production.
    
    \item \textbf{Level 4 - Full MLOps:} Completely automated system with continuous model retraining, verbose centralized metrics, and automatic improvements approaching zero-downtime operations.
\end{itemize}

A recurring tension observed across organizations concerns the balance between use-case driven and framework driven AI adoption. Less mature organizations often achieve quick results by focusing narrowly on individual use cases, but struggle to scale due to fragmented tooling and practices. Conversely, highly centralized organizations may invest heavily in generalized frameworks and governance, yet fail to realize value due to a lack of concrete, operationalized use cases. Organizations that mature successfully tend to iterate between these approaches, allowing early use cases to shape shared frameworks, and evolving frameworks to accelerate subsequent use cases.

\subsubsection{Considerations Based on Maturity Level}
Once you have assessed your organizational maturity, you can use it to guide the implementation of AI application operations and perform adoption using a phased approach. Attempting to implement a full-scale AI operations framework prematurely can lead to inefficiencies and frustration. Instead, organizations should consider the following depending on their organizational ML maturity level:

\begin{itemize}
    \item \textbf{Level 0:} Start with proof of concept projects to build experience, stress-test the organization, and identify gaps. At this stage, it’s crucial to avoid over-engineering and instead prioritize learning and experimentation. You might not reach the application process, but gain organizational experience.
    
     \item \textbf{Level 1:} Go beyond pilot projects, i.e., MVPs and rollouts so that the organization obtains experience of the entire AI application life-cycle. Focus on building core capabilities. This includes establishing basic data governance, reproducible training pipelines, and initial monitoring mechanisms. 
     
    \item \textbf{Level 2:} With automated training in place, organizations can begin to standardize processes and introduce centralized model tracking. This is a good point to start planning for scalable deployment strategies and to define roles and responsibilities for AI operations.
    \item \textbf{Level 3:} Align organizational structures and roles to support AI operations. Organizations can now implement automated deployment pipelines and integrate feedback loops. This enables faster iteration and more reliable performance in production environments. Cross-functional collaboration becomes essential to manage dependencies between the data engineering, data science, ML engineering, and the application teams.
    \item \textbf{Level 4:} Foster a culture of data-driven decision-making and continuous improvement. Here, the focus shifts to feedback mechanisms, optimizing performance, minimizing downtime, and leveraging shared data and models across applications to maximize value.
\end{itemize}

\newpage
\subsection{Roles}
The delineation of roles in AIAppOps intentionally diverges from the more infrastructure-centric characterizations found in major MLOps frameworks (e.g., Google’s \emph{Practitioners Guide to MLOps}~\cite{google2021mlops}) and from the narrow technical abstractions in many industrial “ML team topologies.” In contrast to frameworks that primarily emphasize the automation of pipelines and handoffs between data scientists and ML engineers, the AIAppOps role model explicitly integrates the \emph{application}, \emph{governance}, and \emph{value-realization} perspectives that determine whether machine learning systems sustain value and compliance in production.

Compared to the canonical MLOps lifecycle~\cite{google2021mlops,kreuzberger2023mlops}, where the primary boundary lies between data engineering and ML engineering, we introduce three critical extensions. First, we incorporate \textbf{application engineers}, who translate model capabilities into user-facing and decision-support systems, thus closing the loop from prediction to realized value. Second, we include \textbf{governance and legal functions} as first-class participants, ensuring that compliance, ethics, and accountability are built into daily operations instead of appended as external audits. This prevents what practitioners often call the “compliance-last bottleneck,” where legal review occurs only at deployment time, leading to costly rework or operational risk.
Third, we integrate \textbf{business roles} into the whole process to ensure that value hypotheses, success metrics, and organizational constraints continuously shape technical choices rather than being evaluated post hoc.

A pre-emptive clarification is warranted for readers accustomed to role frameworks such as those in DataOps or DevOps. While these frameworks also assign cross-functional ownership, their boundaries end at data pipelines or software artifacts. AIAppOps extends this logic to the \emph{decision layer}—where data, models, and human oversight intersect—and therefore requires inclusion of roles (e.g., \DME, \LG, \BA, \PM) often treated as peripheral in MLOps literature. By formalizing these as operational participants, the framework reconciles technical excellence with lawful, ethical, and value-aligned operation, in line with the principles of trustworthy AI~\cite{EC_EthicsGuidelines_2019,euai2024}.

Here, we briefly describe the different roles and later map them into the four main processes of AI application operations.

\begin{itemize}
    \item[\DE]\textbf{Data Engineer} Builds and maintains the foundational data pipelines and platforms (like data lakes or feature stores) that make data discoverable, reliable, versioned, and available for consumption.
    \item[\GO]\textbf{Governance Officer / Data Steward} Is responsible for data quality, compliance, and proper data management practices. This includes achieving and maintaining high data readiness (explainable and interpretable data, i.e. information, by man and machine). They define data standards, monitor data usage, and enforce policies related to privacy, security, and regulatory requirements. Their role is crucial in maintaining trust and accountability in data. 
    \item[\DS]\textbf{Data Scientist } Applies statistical analysis, machine learning, and domain knowledge to extract insights from data and data readiness. The insights enable downstream systems and stakeholders to realize value through improved decisions. They need to easily find, understand, and experiment with different datasets and features to build and train models, relying on data catalogs and version control. 
    \item[\MLE]\textbf{Machine Learning  Engineer} Specializes in deploying and maintaining models at scale. They work closely with data scientists to productionize models, optimize performance, and ensure models are robust, reproducible, and integrated into software systems or products.
    \item[\ADE]\textbf{Application Engineers} Designs, develops, and maintains the software and user-facing systems through which AI capabilities create value. Application Engineers encompass software developers, DevOps specialists, and UX or interaction designers working together to integrate AI services into real applications. Their work spans backend and frontend components, APIs, orchestration layers, services, and user interfaces, that connect inference endpoints with other business logic and human workflows. They also manage infrastructure-as-code, automated testing, and CI/CD pipelines to ensure secure, reliable, and continuously evolving operations.
    \item[\BA]\textbf{Business Analyst} Collaborates with stakeholders to understand the business needs and translate them to requirements and success metrics (e.g., KPIs). Works together with data scientists to validate assumptions and interpret model results from a business perspective. Prepares documentation, dashboards, and reports to communicate insights and progress.
    \item[\PM]\textbf{Product Manager} Owns the product vision and ensures that the solution delivers value through defining business requirements, prioritizing features or use cases. Acts as a bridge between technical teams and business stakeholders. The role can also be referred to as product owner when situated closer to the customer.  
    \item[\DME]\textbf{Domain Expert} Brings deep knowledge of the industry or specific business area. They help interpret data in context, validate findings, and ensure that data solutions align with real-world use cases. Their input is essential for data understanding, data curation, accurate modeling and effective decision-making.
    \item[\LG]\textbf{Legal} Ensures that all data practices comply with applicable laws, regulations (e.g., GDPR, HIPAA, CCPA), and contractual obligations. They review data usage policies, assess legal risks, guide consent and data-sharing frameworks, and work with technical and governance teams to mitigate liabilities. Their input is critical in safeguarding the organization from legal and reputational harm.
    
\end{itemize}

In addition to these roles, AIAppOps assumes the presence of business operational roles whose daily work is directly affected by AI-supported or AI-automated decisions. As organizational maturity increases, such roles often evolve into explicitly AI-assisted operational roles, where understanding model behavior, uncertainty, and limitations becomes a natural part of routine work rather than an exception.

\subsubsection{Technical Roles VS Capabilities}

In our stakeholder consortium, we have seen that terms such as \emph{Data Scientist}, \emph{AI Specialist}, or \emph{AI Expert} are used inconsistently across academia, industry, and consulting, often describing vastly different competence profiles under the same label.
In practice, a “Data Scientist” may refer to:
(i) a statistically trained analyst producing decision support for humans,
(ii) an applied machine learning practitioner capable of training deployable models, or
(iii) a broadly trained AI professional with deep understanding of modeling assumptions, uncertainty, system behavior, and operational risk.
These interpretations are summarized and contrasted in Table~\ref{tab:ds_profiles}, together with their implications for AI application operations.

\begin{table}[h]
\centering
\caption{Common interpretations of the “Data Scientist” role and their implications for AI application operations}

\label{tab:ds_profiles}
\begin{tabular}{p{0.25\linewidth} p{0.35\linewidth} p{0.35\linewidth}}
\toprule
\textbf{Data scientist profile} & \textbf{Typical Capabilities} & \textbf{AIAppOps Implications} \\
\midrule
Analytical DS & Statistics, EDA, regression, clustering, dashboards. & Suitable for decision support and early maturity; insufficient for autonomous or regulated AI. \\
Applied ML DS & Supervised/unsupervised ML, feature engineering, evaluation. & Requires ML engineering support; viable for production under controlled scope. \\
AI Systems Expert & Broad AI toolbox, uncertainty, monitoring, system thinking. & Necessary for high-risk, regulated, or autonomous systems. \\
\bottomrule
\end{tabular}
\end{table}

AIAppOps therefore treats role titles as \emph{interfaces}, not guarantees of competence.
What matters operationally is not the title, but the \emph{capabilities actually present in the team} relative to the system’s maturity, criticality, and regulatory exposure.

\subsubsection{Role Competence Misalignment can Lead to Failure}
In our experience, misalignment between perceived and actual competence is a recurring root cause of failed AI projects, particularly when organizations attempt to deploy AI systems beyond decision support into operational or safety-critical contexts without sufficient expertise.
A closely related and equally common failure mode is attempting to solve the wrong kind of problem with the wrong kind of AI.

From applied AI deployments, we have seen that successful initiatives start by identifying so-called low-hanging fruit: problems that are feasible given current data readiness, organizational competence, and the maturity of available methods: \emph{Solve problems that are easy to solve with AI technology and which add meaningful value}.
This does not imply triviality, but appropriateness: selecting problem–method pairs where robustness, evaluation, monitoring, and compliance are achievable from the outset.
Another recurring source of failure in applied AI is the unexamined assumption that machine learning methods are interchangeable. In practice, different problem classes require fundamentally different modeling paradigms, such as probabilistic versus deterministic models, interpretable versus opaque approaches, correlation-based versus causal reasoning, or models that explicitly represent uncertainty rather than point predictions.
The framework therefore treats the ability to select an appropriate modeling paradigm, and to explicitly reject unsuitable ones, as a first-class operational capability. This judgment typically requires access to a broad methodological toolbox and cannot be reliably inferred from role titles alone.

The capability gaps tend to surface most clearly at the boundary between experimentation and operations.
A closely related failure mode is the conflation of demonstrability with deployability. Contemporary tools make it possible to construct visually compelling or functionally impressive AI demonstrations for a wide range of tasks. However, many such systems remain impractical to operate reliably in production due to data dependencies, robustness requirements, latency constraints, organizational processes, or legal and regulatory limitations.
Here, a clear distinction is made between systems that can be demonstrated and systems that can be sustainably operated. Assessing this distinction requires not only implementation skills, but also early judgment of organizational, legal, and lifecycle feasibility—before substantial investment is made. Failure to do so frequently results in “demo success, production failure”, a pattern that is disproportionately costly in regulated or high-stakes environments.

In practice, organizations often mitigate competence gaps by complementing internal teams with external expertise, such as academic collaborators or independent reviewers. AIAppOps treats such second opinions not as a sign of immaturity, but as a risk-management mechanism—particularly when deploying novel methods, operating under regulatory uncertainty, or making irreversible decisions. \emph{What matters is not where expertise resides, but that critical methodological, evaluative, and feasibility judgments are explicitly covered}.

Applied-AI practice consistently distinguishes between specialists who “know their hammer and apply it efficiently” and experts who “understand the toolbox and can effectively choose the right tool for the right problem (and goal).”
The latter capability—method selection under real-world constraints—is critical for identifying low-hanging fruit and avoiding premature commitment to infeasible approaches.

The preceding discussion shows that successful AI application depends not only on technical competence and clear role definitions, but also on the ability to select problems that are feasible given current organizational maturity, methodological readiness, and deployment constraints. In practice, these considerations converge early in the data process, where initial judgments about suitability, risk, and modeling paradigms are formed. However, while data readiness strongly constrains what can be attempted responsibly, it does not by itself confirm that an AI system will deliver sustained value in operation.

From the business perspective, a challenge is that value often becomes completely clear only when candidate models are trained, evaluated, and exposed to realistic operating conditions. As noted by Tiger et al. \cite{tiger2024eva}, assessing whether a task can be solved "to satisfaction" becomes increasingly difficult for complex data and tasks when relying on data inspection alone. It further requires iterative interaction between data understanding, model behavior, and application context~\cite{tiger2024eva}. For this reason, AIAppOps treats the data process as a critical feasibility filter rather than a final arbiter of value. It shapes which hypotheses are worth testing, which modeling paradigms are plausible, and which risks must be monitored downstream.

\subsection{Data process \(\rightarrow\) Data endpoints}
\begin{figure}[ht]
\centering
\includegraphics[width=\linewidth]{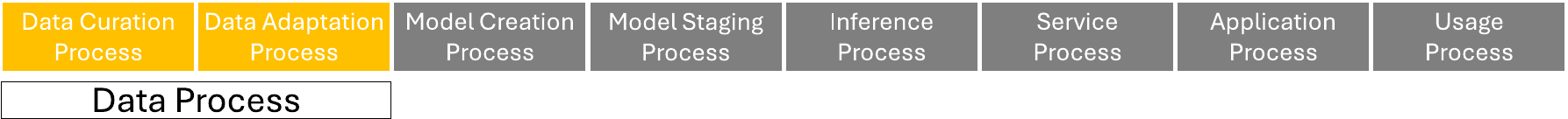}
\end{figure}
The data process begins with sourcing and preparing raw data, converting it into formats usable by machine learning models. This corresponds to raising the dataset’s \emph{data readiness level}~\cite{lawrence2017drl}, from raw and inaccessible data to curated, task-ready datasets. From a business point-of-view, the data process goes further into the business understanding of the existing data and its causality and relevance for the value hypothesis. Deepened understanding of the data from the business perspective can be used to revise the value hypothesis for the downstream processes. In mature organizations, the process goes beyond existing data by also \emph{creating} the data they think they will need for value creation by models.

We divide the process into two sub-processes: \emph{data curation}, which includes model-agnostic steps focusing on data quality, and \emph{data adaptation}, which transforms the data according to modeling assumptions.
The reason for this separation from a process perspective is two-fold. 
First, it clarifies the role separation within the data process where data engineers are primarily involved in data curation, while a data scientist adapts and experiments for model-specific needs.
Second, it highlights the complexity caused by the fact that some data processing is performed closer to, or even during, the model development  (e.g., data augmentation). For explanatory reasons, we have clearly separated the data process from the other processes. However, the data process is commonly an integrated part of the inference and application process since knowledge about the data is often needed in those processes. The data flow can also be circular, i.e., inferences are used as data for other models.

In many AI applications, data now arrives as continuous streams (events, logs, sensor signals) as well as traditional batch extracts. The process therefore accommodates both modalities by enforcing online/offline parity: curation and adaptation steps must be reproducible for batch training and real‑time inference, with versioned data endpoints that include (i) offline snapshots for training and evaluation and (ii) online feature views for low‑latency serving. This parity is essential to avoid training–serving skew, ensure point‑in‑time correctness, and enable consistent monitoring across the lifecycle (cf. \autoref{sec:monitoring}). 

For complex data—including multimodal (text, image, audio, video), temporal (time‑series), and relational/graph‑structured sources, it is often necessary to perform modality‑aware preprocessing, metadata enrichment, and lineage so that downstream modeling assumptions remain valid and auditable. In the end, the process provides versioned data endpoints (feature store and/or stream topics) for model training, inference, and monitoring, with data SLAs that cover both freshness and end‑to‑end latency.

\subsubsection{Data Curation \(\rightarrow\) Clean data}
Data Curation refers to the systematic management and refinement of data to ensure it is fit for purpose. 
It begins with ensuring lawful access to data and continues with structured, automated checks that can run continuously on streams and periodically on batches. The ultimate goal is to produce high‑quality, well‑documented, trustworthy data that is discoverable, reusable, and reproducible.
The key steps involved are:

\begin{itemize}
\item\textbf{Data Accessibility}: The ability to retrieve and legally use the data required to build, train, and deploy a model.
Data can be siloed in different departments where its existence is passed down through "tribal knowledge". To find something, you have to ask multiple people, often leading to dead ends. To mitigate this, the organization can use a centralized searchable repository containing metadata about all available datasets.
Once found, it is necessary to investigate if the data is \textit{allowed} to be used or if there are ways to allow the use of the data. 
It must also be possible to retrieve the data with reasonable efficiency. Accessibility can be hindered by manual handling, e.g., receiving files via email, the size of the data, or the way it is stored.


\item \textbf{Data Understanding} involves both initial investigation of the ability of the accessible data to solve the problem and as well as understanding its quality. This includes both quantitative analysis, e.g., by computing descriptive statistics (mean, median, standard deviation, counts, frequencies etc.) to summarize feature properties, and qualitative analysis, which relies on data visualization (e.g., histograms, scatter plots, box plots) to uncover patterns, distributions, data imbalances, correlations, and potential anomalies. 
For streams, consider rolling/windowed statistics to track distributions, cardinalities, and rare events at scale. For complex data, include modality‑specific descriptors (e.g., language, tokenization stats; image attributes).
This phase is crucial for assessing the data's suitability for the project goals, identifying data quality issues (like missing values or outliers), and forming initial hypotheses that will guide the subsequent data preparation and modeling stages.

\item \textbf{Data Alignment} concerns the explicit alignment between the data used for modeling and the \emph{normative intent} of the application. 
At this stage, teams must decide whether the model should reflect patterns observed in the world as it \emph{is}—including historical biases, structural inequalities, and past decision practices—or the world as it \emph{ought to be}, as defined by legal, ethical, and organizational principles.

Data that faithfully mirrors historical reality can be appropriate and even necessary for certain purposes, such as descriptive analytics, counterfactual reasoning, policy simulation, or synthetic data generation. 
In contrast, decision-support and automated decision-making systems typically require data that is aligned with normative constraints, including non-discrimination, proportionality, and regulatory compliance. 
In such cases, historical biases present in the data must be identified, documented, and actively mitigated rather than passively learned.

Data alignment therefore involves making value-laden choices explicit: selecting target populations, defining protected attributes and acceptable proxies, determining which patterns are permissible to learn, and deciding where rebalancing, reweighting, or data augmentation is required.
These decisions are inseparable from fairness, accountability, and explainability considerations, and must be documented as part of the dataset’s lineage and justification.

\item \textbf{Data Cleaning} involves detecting and resolving errors and inconsistencies in data to enhance its quality. This includes correcting structural errors, standardizing disparate formats, and removing duplicate or irrelevant observations. 
It can also include handling of uncertain data (e.g., through removal or by adding uncertainty descriptions) or missing values (e.g., through imputation or removal), but this can also be addressed in the data adaptation step as it can be model-dependent.
The primary goal is to produce a trustworthy and reliable dataset for robust downstream analysis and modeling.

\item \textbf{Data Validation} ensures that the data follows its assumptions. It often involves programmatic checks the data against a predefined schema and a set of constraints derived from the source, domain knowledge, or a reference dataset. These checks typically include verifying data types and column presence, ensuring values fall within expected ranges, and confirming that key statistical properties (e.g., mean, variance, class distributions) have not significantly deviated. 
The more that is known about the domain, the more precise in-vs-out-of-distribution validation checks can be used.
\end{itemize}

\subsubsection{Data Adaptation \(\rightarrow\) Model-specific data}\label{sec:data-adaption}
Within the model operations pipeline, the Data Adaptation stage is responsible for the transformation of curated data into a specialized format optimized for a target model.  
These transformations can typically not be reused across different types of models or model types.
This generally involves: 
\begin{itemize}
\item \textbf{Data Transformation} within the data curation step performs model-agnostic preprocessing steps that are generally beneficial across a wide variety of machine learning algorithms. 
They address common data quality issues rather than catering to the specific architectural needs of one model.
These transformations are considered "agnostic" because they improve the numerical stability, interpretability, and performance of most models, from simple linear regressions to complex neural networks. A summary of common such transformations is provided in \autoref{tab:data_transformations},
\begin{table}[h!]
\centering
\caption{Summary of common model-agnostic data transformations.}
\label{tab:data_transformations}
\begin{tabular}{p{0.32\linewidth} p{0.35\linewidth} p{0.35\linewidth}}
\toprule
\textbf{Transformation Type}    & \textbf{What It Does}                                  & \textbf{Primary Benefit}                                                                   \\ \midrule
\textbf{Feature Scaling}        & Puts numeric features on a common scale.               & Improves convergence for gradient-based models; essential for distance-based models. \\
\textbf{Categorical Encoding}   & Converts text labels into numbers.                     & Makes data compatible with virtually all ML algorithms.                                    \\
\textbf{Outlier removal} & Removes or caps outliers to reduce their impact on models                 & Improve model performance for common cases.   \\
\textbf{Distributional Transform} & Reduces skewness in data.                              & Stabilizes variance and helps models meet statistical assumptions.                         \\
\textbf{Discretization/binning}         & Groups continuous values into discrete bins.           & Can capture non-linear patterns in a simpler, model-agnostic way.                          \\ 
\textbf{Windowing} & Aggregates over event‑time windows.                              & Encodes dynamics for time‑series/streams. \\ 
\textbf{Embedding Extraction (Text/Image/Audio/Graph)} & Produces vector representations; records model/version.                              & Enable handling of unstructured data; enables multimodal fusion. \\      
\bottomrule
\end{tabular}
\end{table}
    \item \textbf{Model-Specific Feature Engineering} creates or transforms features based on the inductive biases of the model architecture (e.g., tokenization for Transformers, spatial transformations for CNNs). 

    \item \textbf{Labeling} prepares the data for supervised machine learning data points are assigned an informative tag or output value. This label represents the "correct answer" or "ground truth" that the model is expected to predict. The quality, accuracy, and consistency of these labels are paramount, as they directly define the objective function for model training and establish the gold standard for performance evaluation.

    \item \textbf{Data Augmentation} generates variations of the data to enhance training set diversity. Augmentations are often performed during the training process. Here, the versioning can be more related to the types and variability of the augmentations rather than the resulting augmentations. The versioning becomes a trade-off between exact reproducibility and storage or computational resources.
    
\end{itemize}
The output of this stage is a versioned set of data artifacts, managed in a model-specific feature store, which guarantees data lineage and enables the reproduction of training and inference environments.

\subsubsection{Data engineering team \DE~\GO~\DS+(\DME~\LG~\BA)}
This process mainly involve a smaller subset of the team.
Data Engineer and data steward make data discoverable, compliant, high-quality, and versioned. Communication with legal is necessary to clarify uncertainties and obstacles around data usage.
The data scientist is almost always included in this team to explore, validate, and adapt the data (features, labels) together with domain expert(s) to ensure domain fidelity. 
Project manager and business analyst define value hypotheses and data SLAs (freshness, coverage) and align curation/adaptation to downstream needs.

\begin{valueconnection}[Data process]{dataprocesscol}
Data work creates leading indicators for value realization by ensuring that downstream decisions are made on lawful, relevant, timely, and representative data. Typical indicators include dataset coverage and freshness, label quality and agreement, and baseline drift monitors (schema, distribution, and concept proxies). These are tied to application-level hypotheses (e.g., “freshness $<24$h is required to reduce incident handling time by 15\%”), making data Service Level Agreements (SLAs) directly consequential for the value hypothesis.
\end{valueconnection}

\subsection{Model process \(\rightarrow\) Model registry}
\begin{figure}[ht]
\centering
\includegraphics[width=\linewidth]{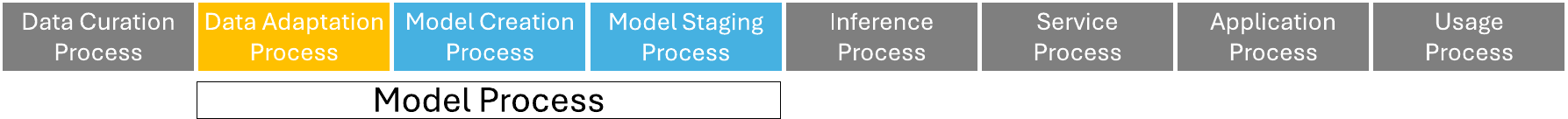}
\end{figure}
The model process transforms curated data into trained and deployable AI artifacts. This phase aligns with the MLOps core pipeline~\cite{kreuzberger2023mlops,google2021mlops}, but it is here extended with governance and staged validation before deployment.
It encompasses the design, training, validation, and operational readiness of models, ensuring that learned representations faithfully encode the intended patterns and behaviors. This process is central to value creation in AI projects and must accommodate experimentation, iteration, and evolution over time. We divide the model process into two sub-processes: \emph{Model creation}, where data becomes a trained model, and \emph{Model staging}, where the model is prepared for real-world use through evaluation and controlled deployment. 

The model process is tightly coupled with the data adaptation process. Since model requirements and architectures evolve throughout experimentation, the features and transformations in data adaptation must remain flexible and co-developed with modeling efforts. This dynamic interplay ensures model fidelity, reproducibility, and performance in context.

\subsubsection{Model Creation \(\rightarrow\) Trained model}
Model creation focuses on converting data into a trained, versioned, and well-documented model artifact. It involves iterative experimentation to identify the most suitable modeling paradigm and configuration, guided by both the structure of the data and the desired application behavior. Crucially, model objectives, loss functions, and evaluation criteria must remain consistent with the data alignment choices established during data curation, ensuring that optimization reinforces the intended normative behavior rather than merely amplifying historical patterns.
Key activities within this phase typically include:

\begin{itemize}
\item \textbf{Model selection} involves choosing a modeling paradigm and concrete architecture that aligns with the structure of the data, the nature of the task, and operational constraints. In organizations with sufficient tooling maturity, automated machine learning (AutoML) techniques may be employed to efficiently explore model classes and hyperparameter spaces, accelerating early-stage experimentation while preserving human oversight over objectives and constraints.

Currently, gradient-boosted decision trees (e.g., XGBoost, LightGBM) are often chosen for structured tabular data, probabilistic machine learning (e.g., Gaussian Processes, causal inference models) for uncertainty- and intervention-aware inference in sensitive and critical domains, and deep learning architectures (e.g., Transformers, CNNs) for unstructured data like text, images, and video. Sequence models (graphical models) such as Conditional Random Fields remain relevant for structured prediction in domains like NLP, automatic control and bioinformatics, while ensemble models like Random Forests offer robustness and interpretability in settings with heterogeneous features. The selection process balances predictive performance with considerations such as training efficiency, interpretability, deployment latency, and maintainability over time.
\item \textbf{Hyperparameter tuning} finds the optimal settings for the training of the selected model or even the model itself~\cite{feurer2015efficient}. Examples of parameters for tuning are learning rates, regularization parameters, or model complexity controls.
In probabilistic machine learning, hyperparameter tuning often overlaps with model selection itself, as adjusting priors or kernel functions effectively changes the model's hypothesis space. Unless the model is very expensive to train, it is often trained during the hyperparameter tuning (the best trained model is picked at the end).
\item \textbf{Pre-training} uses large, general-purpose datasets to initialize model parameters with transferable knowledge, reducing the amount of task-specific data needed later.
\item \textbf{Fine-tuning} specializes a pre-trained model to a specific domain or task using targeted data, often enabling higher performance with less data resource usage. 
\end{itemize}

\textit{Model Creation provides}: A version-controlled model registry that tracks model lineage, training configuration, and evaluation metrics. This enables reproducibility, auditability, and structured experimentation.

\subsubsection{Model Staging \(\rightarrow\) Accessible validated model}

Model staging serves as the bridge between experimental results and operational readiness. It prepares models for deployment through rigorous testing, real-world validation, and alignment with non-functional requirements such as latency, reliability, and compliance.
Accountability for the business outcomes is essential for deciding if a (new) model should be staged. Such decisions are either made with the human-in-the-loop (e.g., the project manager together with support from a data scientist), or automatically based on SLAs.
Typical elements of this sub-process include:

\begin{itemize}
\item \textbf{A/B testing} performs empirical comparison of multiple models—or models against human decision-makers—in live or simulated environments to assess relative performance in business-relevant terms.
\item \textbf{Shadow deployment} deploys a model in parallel with the current production system—without affecting outputs—to gather operational metrics and build trust before full integration.
\item \textbf{Re-training and iteration} update models to account for data drift, concept shifts, or evolving business needs using monitored feedback or scheduled refresh cycles.
\end{itemize}

\textit{Model Staging provides}: A validated deployment pipeline and model delivery framework, ensuring that trained models are production-ready, governed, and performance-assured under real conditions.

\subsubsection{Data science team \DS~+(\MLE~\DME~\PM~\BA)}
Within the model process, the data scientist leads model selection, training, and evaluation; communication with the domain expert ensures valid behavior and failure modes.
The ML engineer codifies experiments and works with the ML operations to ensure reproducibility, packaging, and registry lineage.
The project manager or business analyst are commonly involved in decisions related to bringing models into production to ensure that new models increase the value.

\begin{valueconnection}[Model process]{modelprocesscol}
The model process value connection should answer the question: will the model have good business outcomes? Model metrics are mapped to business outcomes before training. For predictive systems, calibration (how well predicted probabilities match actual outcomes), cost-weighted error (penalizing mistakes based on their business impact), and selective prediction (where the model can abstain or defer decisions when uncertain) are linked to decision cost and safety thresholds. 
For example, in a medical triage system a false negative might be far more costly than a false positive, so error metrics are weighted accordingly. Similarly, abstention is valuable when uncertainty is high, reducing risk.
For generative systems, task success, factuality/groundedness  (accuracy of generated content), refusal and redaction rates (how often the model correctly declines unsafe requests or removes sensitive data), and toxicity/policy conformance are linked to user productivity and risk. 
The model registry stores these mappings with lineage so that value regressions can be traced and explained.
\end{valueconnection}

\subsection{Inference process \(\rightarrow\) Inference endpoint}
\begin{figure}[ht]
\centering
\includegraphics[width=\linewidth]{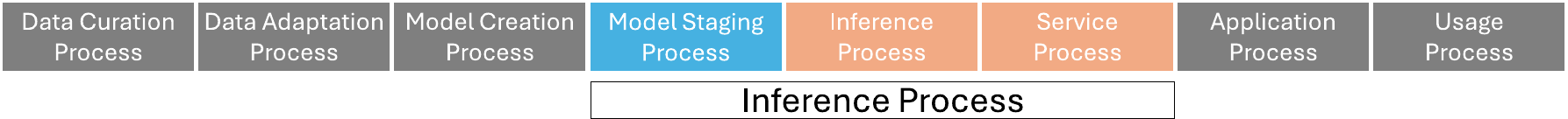}
\end{figure}
The inference process operationalizes trained models by integrating them into real-time or batch-serving environments. This phase transforms models from a theoretical artifact into a system component, reliably executing predictions under performance, availability, and compliance constraints. Industry frameworks typically treat this as the endpoint of MLOps~\cite{google2021mlops}; AIAppOps continues beyond inference to encompass service orchestration, application integration, and user impact evaluation. It includes the infrastructure and logic needed to validate input data, route requests through appropriate inference paths, and optimize throughput. We divide this process into three sub-processes: \emph{Model staging} (covered earlier), \emph{Inference serving}, and \emph{Service orchestration}.

\subsubsection{Inference Serving \(\rightarrow\) Live endpoints}

Inference serving is where trained models are exposed through programmatic endpoints. It ensures that models can be accessed by other systems or users through interfaces optimized for latency, throughput, and security. The inference serving is not necessarily an external endpoint, the serving can also be built into the application if the underlying hardware allows for it.
This sub-process typically involves several key components that ensure the reliability and efficiency of executing model predictions in production environments:

\begin{itemize}
\item \textbf{Input validation}: Ensuring that requests conform to expected schemas and that dangerous or malformed inputs are rejected before reaching the model.
\item \textbf{Input pre-processing}: Perform any necessary data adaptations (\autoref{sec:data-adaption} necessary for the model.
\item \textbf{Batch and real-time handling} for supporting different invocation modes—e.g., batch processing for large datasets, or real-time APIs for instant decisions.
\item \textbf{Scalability} by dynamically allocating resources (e.g., pods, containers, graphical processing units) based on request load using autoscaling policies.
\end{itemize}

\textit{Inference Serving provides}: An inference endpoint with well-defined SLAs, suitable for integration with downstream services. It ensures stability across model calls and can offer  reproducibility, version-handling for API/endpoints as well as observability.

\subsubsection{Service Orchestration \(\rightarrow\) Composable inference}

In some deployments, a single prediction task may require a chain of model invocations, pre/post-processing, or additional business logic. Service orchestration builds such inference graphs—deterministic or agentic—using orchestration frameworks that support robustness, caching, and tracing.
Robust service orchestration is built on a combination of architectural and runtime elements designed to ensure flexibility, control, and resilience:
\begin{itemize}
\item \textbf{Inference pipelines} for sequences of preprocessing, model calls, and postprocessing steps, possibly involving different modalities or models.
\item \textbf{Endpoint selection} chooses which model version or instance to invoke based on metadata such as region, compliance, or A/B configuration.
\item \textbf{Agentic inference orchestration} of LLM agents or graph-based runtime interpreters to decide which submodels to call and in what order, enabling adaptive workflows.
\item \textbf{Resilience strategies}: Circuit breakers, fallbacks, and redundancy mechanisms to guarantee availability and degrade gracefully under fault.
\end{itemize}

\textit{Service Orchestration provides}: A robust service layer on top of inference endpoints that supports composition, fault tolerance, and optimized throughput for diverse applications.

\subsubsection{ML engineering team \MLE~+(\ADE~\PM)}
The main work here is performed by the ML engineer, who provides performant, secure model serving (schemas, autoscaling, version routing). However, they must communicate with the application engineer team to ensure consistency with needs of the application.
On the organizational level the project manager works with SLAs and cost targets.

\begin{valueconnection}[Inference process]{inferenceprocesscol}
Inference SLAs (latency, availability, cost-per-call) are framed as value proxies: if decisions arrive late or are intermittently unavailable, downstream value erodes. Orchestration introduces canaries and A/B configurations so that value hypotheses can be tested safely in production. Resilience strategies (circuit breakers, fallbacks to safer models or retrieval-only answers) are selected to \emph{preserve value under fault} rather than merely maintain uptime.
\end{valueconnection}

\subsection{AI application process \(\rightarrow\) Business-facing services}
\begin{figure}[ht]
\centering
\includegraphics[width=\linewidth]{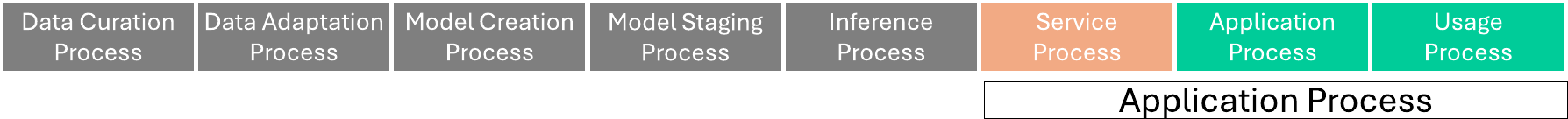}
\end{figure}
Building traditional applications involves a wide range of tasks covered by for example DevOps~\cite{ebert2016devops}. 
Here, we focus on the parts of the application development that are impacted by the use of a data-driven solution. 
The AI application process bridges model outputs with business-facing systems. 
This process ensures that predictions flow into real applications such as dashboards, APIs, or embedded controls—while maintaining CI/CD best practices and aligning with KPIs and business value \& requirements. We divide this process into three sub-processes: \emph{Service deployment}, \emph{Application integration}, and \emph{Usage}.

\subsubsection{Service Deployment \(\rightarrow\) Operational AI services}

This sub-process focuses on deploying and maintaining the logic and infrastructure around AI services. It integrates inference pipelines into backend services that expose predictions via APIs, UIs, or event-driven systems.
To operationalize AI services effectively, this phase integrates infrastructure, automation, and monitoring capabilities through components such as:

\begin{itemize}
\item \textbf{DevOps for AI services} builds and maintains CI/CD pipelines that validate and deploy service logic, infrastructure as code for reproducibility, and canary or shadow deployments for gradual rollout.
\item \textbf{Monitoring and alerting}: Continuous observation of latency, throughput, error rates, and business metrics.
\item \textbf{Compliance pipelines}: Automated checks to ensure that services using AI maintain required documentation, audit logs, and approval checkpoints.
\end{itemize}

\textit{Service Deployment provides}: An AI-powered service layer with observable, maintainable, and policy-compliant interfaces.

\subsubsection{Application Integration \(\rightarrow\) System-level access}

Once the AI service is deployed and observable, it must be woven into the broader software and organizational ecosystem. This integration ensures that predictions reach users or systems at the right time, in the right format, and with the necessary context. Whether the service is consumed internally or externally, this phase is where AI begins to directly shape workflows, decisions, and products.
Key aspects of integration typically include:
\begin{itemize}
\item \textbf{Interface design and exposure} through structuring the service contract — what inputs are accepted, what outputs are returned, and under what guarantees (e.g., latency, reliability, interpretability). Interfaces must remain stable and versioned over time to support downstream compatibility.
\item \textbf{Embedding in applications} by connecting AI services to operational systems such as dashboards, case management tools, process automation engines, or end-user applications. This often requires adapting output formats, handling errors gracefully, and aligning responses with business needs.
\item \textbf{Security and access governance} controls who can invoke AI services, how data flows are logged and monitored, and whether outputs meet domain-specific compliance requirements (e.g., access audits, rate limiting, encryption at rest and in transit).
\end{itemize}

\textit{Application Integration provides}: An application that makes AI functionality available to users, or automated agents.

\subsubsection{Usage \(\rightarrow\) Real-world impact}

This final stage evaluates how well the deployed application supports real-world tasks and decision-making. It focuses on measuring outcomes, learning from user behavior, and closing the loop between model predictions and business results.
To close the loop between prediction and business value, this phase emphasizes mechanisms for impact measurement and user-driven feedback:

\begin{itemize}
\item \textbf{Task performance analysis} measures whether model outputs result in improved efficiency, quality, or user satisfaction.
\item \textbf{User feedback loops} captures corrections, exceptions, or suggestions from human users and feeding them back into training or retraining loops.
\item \textbf{Key Performance Indicator (KPI) alignment} correlates application behavior with key business metrics (e.g., cost savings, error reduction, time-to-resolution).
\end{itemize}

These signals reveal whether the AI system is delivering its intended value and operating safely in context. If deviations occur, they indicate value drift, i.e., a gap between technical performance and business outcomes.
These deviations must feed back into upstream processes. For example, frequent overrides or deferrals may indicate gaps in training data or calibration issues, prompting adjustments in the Data Process (e.g., resampling underrepresented cohorts) or Model Process (e.g., retraining with updated thresholds or fairness constraints). Latency-related usage problems can trigger changes in the Inference Process, such as autoscaling or routing optimizations, while usability or trust issues may require refinements in the Application Process, including interface design and explanation mechanisms. By closing this loop, organizations ensure that operational insights continuously shape data, models, and services, preserving alignment with business goals and regulatory requirements.
Thus, the usage process is one of the key sources of information in monitoring process (\autoref{sec:monitoring}), which ties together the business value with the four main processes.  

\textit{Usage provides}: Real-world impact validation and continuous signals for improving models, services, or interfaces.

\subsubsection{Core application team \ADE~\DME~\BA~\PM~\GO~\LG}
Here, the main work is performed by the application engineers to develop the AI-backed services and integrate inference into systems and UIs.
However, this process may involve many of the roles that have been part of the project as both the data processing and model creation impact the usage within the application. Thus, it is important to either provide necessary documentation and/or allow for communication with the team members involved in the upstream tasks of processing the data and developing the models.
The business analyst's role here is to finally measure the impact, build dashboards for monitoring metrics, and together with the team analyze discrepancies or misalignments that should be fed to the earlier parts of the process (e.g., update requirments, data processing, model metrics etc.). 
It is also important to involve business operations as a functional application may still fail if the business operations is not ready to use it. 
A new application may for example involve new ways of working, e.g. AI assisted, which is part of the business operations rather than the application development.
The project manager steers the roadmap via KPIs.

\begin{valueconnection}[Application process]{applicationprocesscol}
The application is where value is realized and observed. Core signals include adoption and override rates, time-in-state and rework, error and escalation patterns, and user satisfaction. These are wired back to inference, model, and data as structured feedback (labels, prompts/policies, features), enabling targeted improvements that move the same business KPIs that justified the project.
\end{valueconnection}

\subsection{Monitoring \(\rightarrow\) Value-drift tracking}
\label{sec:monitoring}
Monitoring machine learning systems has long been recognized as a complex, multi-dimensional challenge~\cite{naveed2025monitoring}.
Organizations often struggle to synthesize signals from data, models, infrastructure, and applications into a coherent operational picture.
In AIAppOps, this challenge is addressed by integrating the monitoring of data drift, model behavior, and application-level KPIs into a unified feedback loop that ties technical observability to business outcomes.
Through this integration, monitoring becomes not merely a safeguard, but a continuous process for ensuring that deployed AI systems remain reliable, safe, lawful, and valuable throughout their lifetime.
Whereas traditional MLOps frameworks typically emphasize technical metrics such as model accuracy, latency, or data drift~\cite{google2021mlops,kreuzberger2023mlops}, 
AIAppOps extends the scope of monitoring into a continuous socio-technical activity that links \emph{technical observability} to \emph{business value}, \emph{human oversight}, and \emph{compliance}. 
It turns telemetry into assurance by embedding measurement, explanation, and adaptation across all life-cycle stages.

\subsubsection{Purpose and scope}

Monitoring in AIAppOps pursues four intertwined goals.
First, it safeguards the \textbf{effectiveness} of predictive and generative systems by tracking performance, calibration, and alignment over time.
Second, it maintains \textbf{service quality} by ensuring that availability, latency, and cost adhere to agreed Service-Level Objectives (SLOs).
Third, it reinforces \textbf{trustworthiness} through continuous checks on safety, fairness, privacy, and explainability under evolving conditions.
Finally, it secures \textbf{value realization} by verifying that technical outcomes continue to produce the intended organizational or societal impact.

Recent reviews highlight that industrial monitoring often focuses narrowly on technical metrics, leaving fairness, interpretability, and business impact under-represented~\cite{naveed2025monitoring}. 
AIAppOps explicitly closes this gap by unifying these perspectives within one feedback loop.

\subsubsection{Layers of observability: what to monitor}

AIAppOps defines monitoring as a layered system of observability with each layer corresponding to a critical dimension of the AI lifecycle:

\begin{itemize}
\item \textbf{Data:} schema adherence, completeness, (class) balance, distributional and concept drift, and freshness; derived from data readiness concepts~\cite{lawrence2017drl,tiger2024eva}.
\item \textbf{Model:} performance, calibration, uncertainty, out-of-distribution (OOD) behavior, and fairness.
\item \textbf{Inference and infrastructure:} latency, throughput, error rates, resource use, and resilience.
\item \textbf{Pipeline:} CI/CD health, reproducibility, and lineage consistency.
\item \textbf{Governance:} version-to-service mapping, access control, consent, and auditability in accordance with ISO/IEC 42001:2023 and AI Act Article 72~\cite{iso42001,euai2024}.
\item \textbf{Application and value:} user feedback, adoption, override rates, KPI impact, and cohort equity.
\end{itemize}

Each layer contributes to validating the system’s declared \emph{value hypothesis}. 
Importantly, monitoring must distinguish between drift in observed reality and drift away from the system’s intended normative alignment, as improvements in predictive accuracy may still degrade fairness, legality, or public trust.
A deviation in any layer can erode value, trust, or compliance, motivating corrective action through the ``Decide \& Adapt'' loop.

\subsubsection{Dimensions and automation}

Monitoring in AIAppOps operates along three main dimensions.
The \emph{purpose dimension} concerns quality, robustness, and value alignment.
The \emph{temporal dimension} distinguishes proactive early-warning signals, reactive incident detection, and retrospective analysis for learning.
The \emph{automation dimension} defines how these observations are handled, ranging from manual reviews to fully automated responses (system detects an issue and takes corrective action without human intervention).
For example, when the monitoring system detects data drift beyond a predefined threshold, it automatically triggers a model retraining pipeline and deploys the updated model through CI/CD. Similarly, if inference latency exceeds the Service Level Objective (SLO), the system autoscales resources or reroute traffic to a backup model immediately. 
In practice, it is often good to favor semi-automated setups that combine automatic detection with human judgment, balancing efficiency and accountability~\cite{naveed2025monitoring}.

\subsubsection{Data and architecture for monitoring}

Effective monitoring depends on the systematic collection and linkage of telemetry across data, model, and inference pipelines.
A robust monitoring architecture collects, links, and stores telemetry across data, model, and inference pipelines. 
Two principles are central:  
(i) \emph{Feature and prediction logging} at inference time, including model and data version identifiers, to enable later comparison with observed outcomes; and  
(ii) \emph{Lineage tracking} to guarantee that every signal can be traced to its origin and context.  
Such joinable records underpin reproducibility, auditability, and compliance, without reference to specific platforms,
and align with best practices for structured validation and production-readiness testing in machine learning pipelines~\cite{breck2017mltest,odena2019tensorflow}.

\subsubsection{Methods and metrics}

\paragraph{Statistical and ML-based monitoring.}
Statistical and machine-learning–based monitoring relies on metrics such as drift detection (data and concept), calibration error, uncertainty quantification, and selective prediction or abstention~\cite{guo2017calibration,hendrycks2016baseline,lakshminarayanan2017deep}.
In predictive systems, these indicators preserve decision quality and safety under shifting data distributions.
In generative systems, additional monitors assess groundedness, citation fidelity, toxicity and policy conformance, refusal and redaction behavior, and retrieval quality in retrieval-augmented generation (RAG).
Together, these measures ensure that model outputs remain factual, appropriate, and aligned with both technical and organizational policies.

\paragraph{Logical and formal monitoring.}
Quantitative metrics alone cannot guarantee safety, legality, or policy compliance.
Formal methods extend monitoring into the logical domain, enabling the specification and continuous verification of properties such as
“the probability of violating constraint $X$ within 5 seconds must remain below 1\%”.
Foundational research in runtime verification has established the theoretical and algorithmic basis for such guarantees~\cite{havelund2018runtime,falcone2019tutorial}.
In cyber–physical and autonomous systems, temporal logic monitoring—particularly Signal Temporal Logic (STL)—has emerged as a core mechanism for specifying and measuring dynamic behaviors~\cite{donze2013efficient,deshmukh2015robustonline,mitsch2018verified,jaksic2018algebraic}.
These works introduced notions such as \emph{robust satisfaction}, \emph{partial monitoring}, and \emph{algebraic composability}, which together make formal assurance computationally tractable in real time.
Recent extensions further address uncertainty and distributional shift, combining probabilistic reasoning with runtime prediction~\cite{zhao2023robuststl}.

Probabilistic Signal Temporal Logic (ProbSTL)~\cite{tiger2020probstl} exemplifies this line of work by integrating probabilistic uncertainty with temporal constraints and providing incremental algorithms for real-time evaluation.
Developed within the context of safety-aware autonomous systems~\cite{tiger2023safetyaware}, ProbSTL demonstrates how runtime monitoring can incorporate both stochastic predictions and confidence measures, supporting adaptive, self-assessing AI behavior.
Within AIAppOps, these formalisms represent the highest maturity level of monitoring—\emph{runtime verification under uncertainty}—where empirical observability is complemented by mathematically grounded assurance~\cite{varshney2017safety,zhang2020formal,amodei2016concrete}.

\subsubsection{Experimentation-aware releases and response}

Monitoring and experimentation are coupled.  
New models or configurations are deployed through canary or shadow deployments, where monitored metrics act as gates for promotion or rollback~\cite{google2021mlops}.  
Detected degradations activate structured \emph{runbooks}—predefined operational guides that specify ownership, assessment steps, and fallback or rollback actions—an established practice in Site Reliability Engineering~\cite{beyer2016sre,ebert2016devops}.
Within MLOps, such runbooks operationalize incident response and ensure consistent, accountable decision-making when monitored indicators deviate~\cite{naveed2025monitoring,google2021mlops}.
Each response is logged and linked to its originating signal, ensuring traceability and learning for future iterations.

\subsubsection{Governance, compliance, and human oversight}

Monitoring also serves as the practical mechanism for fulfilling post-market surveillance and documentation obligations.
The EU AI Act requires high-risk systems to maintain a post-deployment monitoring plan that continuously assesses performance, safety, and compliance indicators~\cite{euai2024}.
International standards such as ISO/IEC 23894:2023 and 42001:2023 reinforce this principle by mandating continuous improvement cycles.
In AIAppOps, dashboards are designed not only for observability but for \emph{interpretability}—presenting uncertainty visualizations and explanatory cues that allow human operators to calibrate trust and intervene when necessary.
In this way, AIAppOps generalizes assurance principles from system-level runtime verification~\cite{tiger2023safetyaware} to organizational oversight, ensuring that compliance and human judgment remain anchored in verifiable evidence.
This approach is consistent with broader AI assurance frameworks~\cite{varshney2017safety,doshi2017towards,bussone2015explanations},
linking continuous monitoring with calibrated human trust and accountable oversight.

\subsubsection{Maturity roadmap}

Monitoring capability evolves with organizational maturity:
\begin{itemize}
    \item \textbf{Levels 0–1:} Ad-hoc checks; manual dashboards for latency and availability.
    \item \textbf{Level 2:} Centralized metrics; data-drift detection and feature/prediction logging.
    \item \textbf{Level 3:} Automated alerting, calibration and OOD monitors, A/B testing, and retrain triggers.
    \item \textbf{Level 4:} Unified dashboards integrating business KPIs, fairness and compliance indicators, and formal runtime verification.
    \item \textbf{Level 5:} Anticipatory; predicting value outcomes getting compromised before it happens and proactively engaging in activities to keep the organization and its systems value-aligned.
\end{itemize}
Advancing along these maturity levels should follow demonstrated operational need and organizational competence.
Premature automation risks obscuring accountability and undermining trust, which are two of the outcomes monitoring is designed to preserve.
For example,  when responses to monitoring signals (e.g., retraining, rollback, scaling) are fully automated without clear ownership or documented decision criteria, it becomes difficult to trace who is responsible for the change and why it happened. This lack of transparency can lead to obscured accountability.
Also,  if automated actions occur without visibility or justification, stakeholders lose confidence in the system’s reliability and governance. Stakeholders may perceive the AI as a "black box" making uncontrolled changes, which erodes trust.

The signals and decisions established through monitoring directly feed the \emph{Observe} and \emph{Decide \& Adapt} activities described in Section~\ref{sec:value-driven-operations}, ensuring that every observed deviation—technical or organizational—is tied to measurable value outcomes.

\begin{valueconnection}[Monitoring]{monitoring}
Monitoring transforms observability into value control. 
Data and model signals (drift, calibration, OOD) anticipate value loss before KPIs move; 
inference SLOs safeguard timeliness and user experience; 
application metrics and user feedback close the loop on realized outcomes. 
Feature- and prediction-level lineage enables precise attribution when value drifts, 
focusing improvements where they yield the highest return~\cite{google2021mlops,naveed2025monitoring}.
\end{valueconnection}

\subsection{Assess, Analyze, Improve}
Continuous assessment within AIAppOps extends beyond technical metrics to a full socio-technical learning process that links operational signals to organizational improvement.
Every observed deviation—be it technical, procedural, or ethical—initiates a structured cycle of assessment, analysis, and improvement.

\textbf{Assessment} refers to the continuous evaluation of monitored signals across the data, model, inference, and application layers. It determines whether observed deviations affect the declared value hypothesis or trustworthiness requirements. Examples include drift in data distributions, loss of calibration, or shifts in user behavior.

\textbf{Analysis} focuses on understanding root causes and their implications. Technical causes (e.g., concept drift, data imbalance, or model overfitting) are linked to organizational and process factors (e.g., missing validation steps, insufficient human oversight, or inadequate documentation). This diagnostic step is crucial for preventing repeated errors and for turning incidents into learning opportunities.

\textbf{Improvement} transforms insights into structured change. This may include both technical changes, such as retraining models, updating labeling practices, improving monitoring thresholds, refining user interfaces, or adjusting governance rules, as well as organizational such as adjustments to governance rules and decision policies. Each improvement is logged, versioned, and validated, ensuring reproducibility and accountability. 

\noindent Over time, this cyclical process increases both \textit{AI maturity} and \textit{organizational resilience}. It evolves from reactive error correction to proactive optimization and anticipatory governance. By explicitly linking technical evidence to human decision-making and value outcomes, AIAppOps ensures that improvement efforts are data-driven, ethical, and aligned with organizational goals.

\subsection{Trustworthy AI and Compliance}
Trustworthy AI in the European context is built upon three mutually reinforcing pillars: \textbf{lawfulness}, \textbf{ethical alignment}, and \textbf{technical robustness}. These pillars, defined in the European Commission’s \emph{Ethics Guidelines for Trustworthy AI} \cite{EC_EthicsGuidelines_2019} and reinforced by the \emph{EU Artificial Intelligence Act} \cite{euai2024}, establish a foundation for responsible AI development and deployment. AIAppOps operationalizes these principles throughout the AI lifecycle by embedding them in its processes for data governance, modeling, monitoring, and value-driven improvement.

Trustworthy AI is not a static certification but a continuous state achieved through deliberate monitoring, documentation, and adaptation. In AIAppOps, these requirements are mapped to operational mechanisms as follows:

\definecolor{trustlight}{HTML}{E6EEF8}

\newtcolorbox{trustitem}{
  enhanced,
  colback=trustlight,
  colframe=white,
  boxrule=0pt,
  arc=1mm,
  left=4mm,right=4mm,top=2mm,bottom=2mm,
  before skip=4pt,after skip=6pt
}

\begin{trustitem}
\textbf{1.~Human Agency and Oversight}\\
AIAppOps establishes \emph{human-in-the-loop (HITL)} and \emph{human-on-the-loop (HOTL)} modes to preserve situational awareness and control, in line with EU AI Act Article~14. Decision deferral mechanisms and calibrated uncertainty visualizations empower operators to interpret and override outputs. Dashboards fuse reliability, explainability, and confidence signals to support informed human judgment.
\end{trustitem}

\begin{trustitem}
\textbf{2.~Technical Robustness and Safety}\\
Reliability under uncertainty is assured through \emph{probabilistic runtime verification} (ProbSTL), calibration and out-of-distribution monitoring, adversarial robustness testing, and reproducible builds. Fail-safe fallback and recovery policies sustain traceability and operational continuity during abnormal behavior.
\end{trustitem}

\begin{trustitem}
\textbf{3.~Privacy and Data Governance}\\
Data pipelines integrate GDPR-aligned lineage tracking, retention, and consent management (AI Act Article~10). Automated minimization, pseudonymization, and masking are enforced within CI/CD flows, while access logging and accountability policies ensure lawful, transparent data use.
\end{trustitem}

\begin{trustitem}
\textbf{4.~Transparency}\\
Standardized documentation—\emph{Data Sheets}, \emph{Model Cards}, and \emph{Decision Logs}—make datasets, models, and decisions auditable (AI Act Article~13). Versioned lineage provides traceable insight into system assumptions, capabilities, and limitations across releases.
\end{trustitem}

\begin{trustitem}
\textbf{5.~Diversity, Non-discrimination, and Fairness}\\
Bias detection, mitigation, and accessibility testing are embedded across data curation, model evaluation, and monitoring. 
Their meaning depends on the declared data alignment intent: whether the system describes historical reality or enforces normative decision principles.
Diverse datasets and user cohort testing ensure equitable performance and fair treatment across demographics and contexts.
\end{trustitem}

\begin{trustitem}
\textbf{6.~Societal and Environmental Well-being}\\
AIAppOps tracks environmental impact (energy per inference, training footprint) and relates system performance to societal indicators. This embeds sustainability and democratic accountability within operational metrics, ensuring AI contributes positively to long-term societal resilience.
\end{trustitem}

\begin{trustitem}
\textbf{7.~Accountability}\\
Clear ownership of artifacts, roles, and lifecycle stages ensures responsibility. Audit logs, compliance dashboards, and ISO/IEC~42001:2023 alignment provide evidence of continuous stewardship and regulatory conformity throughout operation.
\end{trustitem}

By aligning process design and monitoring infrastructure with these seven requirements, AIAppOps transforms abstract ethical principles into verifiable operational practices. 
Compliance is not treated as an external audit phase, but as an \emph{embedded operational property}—maintained through continuous feedback, documented evidence, and multi-role accountability. 

At higher organizational maturity, trustworthy AI becomes inseparable from operational excellence: systems are not only safe and lawful, but also demonstrably aligned with human values, societal benefit, and long-term sustainability.

\definecolor{trustcol}{HTML}{005F9E} 
\newtcolorbox{trustbox}[1][]{
  colback=trustcol!4!white,
  colframe=trustcol!70!black,
  coltitle=black,
  fonttitle=\large\bfseries\sffamily,
  fontupper=\sffamily,
  boxrule=0.3mm,
  arc=1mm,
  toptitle=3mm,
  bottomtitle=2mm,
  title=#1,
  left=6mm,
  right=6mm,
  bottom=2mm,
  top=2mm,
  borderline west={2pt}{0pt}{trustcol!60!black},
  enhanced
}

\begin{trustbox}[colback=white!97!gray,colframe=black!15,arc=0.5mm,boxrule=0.3pt,
title=\textbf{Lawful, Ethical, and Robust AI — The Foundations of Trustworthy AI}]
\begin{itemize}
\item \textbf{Lawful AI:} Complies with the EU AI Act, GDPR, and ISO/IEC~42001. Ensures documentation, human oversight, and accountability. 
\item \textbf{Ethical AI:} Promotes fairness, transparency, and human agency, avoiding bias and misuse. 
\item \textbf{Robust AI:} Maintains safety, resilience, and explainability under uncertainty through probabilistic and formal monitoring. 
\end{itemize}
\noindent Together, these principles define the socio-technical baseline that AIAppOps operationalizes through its continuous lifecycle.
\end{trustbox}

\section{Value-driven Operations}
\label{sec:value-driven-operations}
This section defines the operational loop that links technical observability to business outcomes. It consists of four recurring activities—\emph{Hypothesize, Instrument, Observe, Decide \& Adapt}—executed at increasing scope as systems mature (use case $\rightarrow$ product $\rightarrow$ portfolio).

\subsection{Hypothesize}
Teams articulate a value hypothesis that specifies (i) intended outcomes and beneficiaries, (ii) measurable proxies and decision thresholds, and (iii) constraints and guardrails.
\begin{itemize}
  \item \textbf{Outcomes:} financial (cost/revenue), operational (throughput, latency, quality), risk/compliance (incidents avoided, audit timeliness), and public value (access, equity, satisfaction).
  \item \textbf{Proxies and thresholds:} early-stage technical metrics that predict target outcomes (e.g., calibration error $\rightarrow$ triage safety; groundedness $\rightarrow$ casework quality) with pre-agreed thresholds that gate POC$\rightarrow$MVP$\rightarrow$rollout.
  \item \textbf{Guardrails:} safety, privacy, policy, and legal requirements that must hold at each stage (e.g., data minimization, audit trails, transparency obligations).
\end{itemize}

\subsection{Instrument}
Before building, teams ensure that value is \emph{measurable} and \emph{traceable}.
\begin{itemize}
  \item \textbf{Telemetry plan:} event schemas for task outcomes, human overrides/deferrals, explanation/justification capture, and user feedback; cohort and feature flags for experiments.
  \item \textbf{Evaluation harness:} offline test suites (predictive: cost-sensitive, calibration, drift; generative: task/eval sets, groundedness, refusal/policy tests) aligned to the hypothesis.
  \item \textbf{Lineage and auditability:} versioning for data, code, models, prompts/policies, and configurations; reproducible builds; immutable logs that link predictions/generations to inputs and model versions.
  \item \textbf{Privacy and compliance hooks:} access controls, personally identifiable information handling, consent/accountability records, and documentation artifacts integrated into CI/CD.
\end{itemize}

\subsection{Observe}
Run staged experiments and operations with decision-quality visibility.
\begin{itemize}
  \item \textbf{Staged delivery:} POC in a sandbox or shadow deployment; MVP to limited cohorts; progressive rollout with canarying and A/B.
  \item \textbf{Dashboards:} joint views that bind technical signals (latency, error, drift, hallucination/failure modes) to business KPIs (e.g., handle time, deflection, risk events).
  \item \textbf{Degradation and drift:} SLO/SLA alerts tied to value impact; statistical and semantic drift detectors that trigger upstream investigation.
\end{itemize}

\subsection{Decide \& Adapt}
When signals deviate (positively or negatively), changes are targeted where they have the highest value leverage.
\begin{itemize}
  \item \textbf{Data:} re-sample cohorts, re-label, close coverage gaps, update retention/freshness SLAs.
  \item \textbf{Model:} adjust loss/thresholds and/or hyper parameters, retrain or fine-tune, introduce abstention or selective routing, expand evaluation suites.
  \item \textbf{Inference:} change routing policies, enable fallbacks (e.g., retrieval-only answers), adjust autoscaling and caching to hit value-critical SLOs.
  \item \textbf{Application:} modify UX and integration points, change human-in-the-loop placement, evolve policy prompts and explanations.
\end{itemize}
Decisions and their rationale are recorded with links to metrics, artifacts, and approvals to maintain traceability.

\subsection{Operating cadence and ownership}
Value is a product property with clear ownership.
\begin{itemize}
  \item \textbf{Roles:} a business owner (accountable for outcomes) and a technical owner (accountable for service and model behavior) jointly steward the value hypothesis.
  \item \textbf{Cadence:} frequent (e.g., weekly) operational reviews of SLOs and safety signals; regular  (e.g., monthly) value reviews of KPI movement and cohort equity; regular (e.g., quarterly) portfolio reprioritization based on realized value and risk.
  \item \textbf{Change management:} pre-commit rollback criteria; maintain model/prompt/policy change logs; ensure documentation (model cards, data sheets, decision records) is updated as part of CI/CD.
\end{itemize}
This cadence closes the loop between observed impact and technical change, ensuring that deployed AI systems remain safe, effective, and valuable over time.
Where the monitoring layer provides probabilistic and formal assurance of technical behavior, the value-driven loop ensures that these assurances translate into measurable organizational and societal outcomes.

\section{Conclusion and Outlook}
AI Application Operations (AIAppOps) establishes a socio-technical framework that extends traditional MLOps to encompass the full lifecycle of AI systems—from initial value hypothesis through data, model, inference, and application, to post-deployment monitoring and improvement. 
Its defining feature is the continuous feedback between these processes, guided by a value-driven meta-process that ensures every technical activity contributes to measurable and sustainable outcomes.

By embedding observability, uncertainty awareness, and human oversight into all lifecycle stages, AIAppOps bridges the gap between engineering practice and trustworthy AI principles. 
It aligns with the European Union’s vision of \emph{lawful, ethical, and robust} AI as described in the \emph{Ethics Guidelines for Trustworthy AI} and the \emph{EU AI Act}, operationalizing their requirements through concrete, monitorable processes.

The framework is intentionally modular and adaptable: it can be implemented incrementally according to organizational maturity, domain criticality, and regulatory obligations. 
Organizations at early maturity levels can begin with data readiness and monitoring foundations, while mature AI programs can integrate formal verification, probabilistic assurance, and full governance alignment.

\textbf{Looking forward}, several directions emerge for advancing both research and practice:
\begin{itemize}
\item \emph{Standardization:} Convergence between AIAppOps and ISO/IEC~42001, 23894, and future AI assurance standards can formalize the operational backbone of trustworthy AI. 
\item \emph{Automation:} Integrating runtime reasoning and formal verification (e.g., ProbSTL) with adaptive retraining loops can yield self-assessing AI systems that maintain safety and alignment autonomously.
\item \emph{Open ecosystems:} Shared reference implementations, open-source observability stacks, and benchmarking of AI monitoring architectures will accelerate cross-sector adoption.
\item \emph{Human oversight at scale:} Designing interfaces that communicate uncertainty, provenance, and rationale effectively to human operators remains a frontier for socio-technical AI design.
\end{itemize}

AIAppOps thus provides not only a practical framework for organizations today but also a foundation for tomorrow’s AI governance infrastructures, where value, trust, and accountability evolve together through evidence-based operations.

\section*{Acknowledgements}
This work was financed by VINNOVA and the organization participating in the Data-Driven Organizations (DDO) project.
The participating organizations were Aixia, Hewlett Packard Enterprise, Hopsworks, IBM, Linköping University, NetApp, Predli, Proact, RISE, RedHat, Region Halland, Sahlgrenska University Hospital, Statistics Sweden (Statistiska Centralbyrån), The Swedish Tax Agency (Skatteverket), Stormgrid, The Swedish Transport Administration (Trafikverket), Volvo Parts, Region Västra Götaland, Santa Anna, and AI Sweden.

\newpage
\section*{Glossary of Key Terms and Concepts}
\label{sec:glossary}

\glossentry{Abstention}
{The ability of a model to decline making a prediction when confidence is low.}
{A medical diagnosis model might abstain when uncertainty exceeds a threshold, triggering human review.}
{Abstention improves safety and trustworthiness, especially in high-stakes domains.}

\glossentry{Bias and Fairness}
{Bias refers to systematic errors that unfairly disadvantage certain groups; fairness refers to mitigation strategies and equitable outcomes.}
{A hiring model that under-selects women due to biased training data exhibits gender bias.}
{Addressing bias is critical for legal compliance, ethical responsibility, and public trust.}

\glossentry{Calibration}
{The alignment between a model’s predicted probabilities and actual outcomes.}
{A calibrated model predicting 0.8 probability of churn will see roughly 80\% of those cases churn in reality.}
{Calibration ensures that confidence scores are meaningful, supporting risk-sensitive decisions and uncertainty estimation.}

\glossentry{Canary Deployment}
{A deployment strategy where a new version of a model or service is rolled out to a small subset of users or requests before full release.}
{Only 5\% of traffic is routed to a new model initially, allowing engineers to monitor its behavior before scaling up.}
{Canary deployments reduce risk by detecting performance or compliance issues early in the release process.}

\glossentry{CI/CD (Continuous Integration / Continuous Deployment)}
{A set of software engineering practices that automate the building, testing, and deployment of code and models into production.}
{A CI/CD pipeline might automatically retrain a model when new data arrives, test it against validation criteria, and deploy it to a staging environment.}
{CI/CD accelerates iteration, reduces human error, and ensures that models and services can evolve continuously without sacrificing reliability.}

\glossentry{Concept Drift}
{A change in the relationship between input features and the target variable over time.}
{A spam classifier may degrade as spammers adopt new tactics that change the underlying concept of “spam.”}
{Concept drift requires model retraining or adaptation to maintain predictive performance.}

\glossentry{Containerization}
{Packaging software and its dependencies into lightweight, portable containers to ensure consistent execution across environments.}
{A model inference service packaged in a Docker container behaves identically in development, testing, and production environments.}
{Containerization simplifies deployment, scaling, and reproducibility, which are essential for reliable AI operations.}

\glossentry{Data Catalog}
{A searchable inventory of datasets and metadata, often enriched with lineage, quality metrics, and ownership information.}
{A catalog entry might describe a customer dataset, its schema, last update, and associated access policies.}
{Data catalogs make data discoverable and governable, accelerating development and reducing duplication.}

\glossentry{Data Drift}
{A change in the statistical properties of input data over time, which can degrade model performance.}
{A credit scoring model trained on pre-pandemic data may become inaccurate as customer behavior shifts post-pandemic.}
{Detecting and responding to data drift is crucial for maintaining model accuracy and reliability in production.}

\glossentry{Data Governance}
{The set of policies, roles, and processes that ensure data is used responsibly, securely, and in compliance with regulations.}
{A governance framework may define data access controls, retention policies, and consent management procedures.}
{Strong governance underpins trust, legal compliance, and operational reliability.}

\glossentry{Data Sheet}
{A standardized documentation format describing a dataset’s origin, composition, collection process, and potential risks.}
{A medical dataset’s data sheet might document patient consent procedures and known sampling biases.}
{Data sheets help assess suitability, legal compliance, and ethical risks associated with data use.}

\glossentry{Explainability}
{Techniques and tools that make model decisions understandable to humans.}
{SHAP values might explain which features, for a specific learned model, most influenced a loan approval decision.}
{Explainability supports trust, debugging, accountability, and compliance with regulations like the AI Act.}

\glossentry{Feature Store}
{A centralized repository for storing, managing, and serving machine learning features consistently across training and inference.}
{A feature store might provide preprocessed customer features to both a churn prediction model and a recommendation engine.}
{Feature stores improve consistency, reuse, and governance of data pipelines, accelerating development and reducing errors.}

\glossentry{Feedback Loop}
{A mechanism by which model predictions and user interactions are fed back into training or evaluation processes.}
{User corrections to an AI assistant’s answers can be used to improve future model performance.}
{Feedback loops enable continuous improvement and adaptation to real-world conditions.}

\glossentry{Hallucination}
{The generation of plausible-sounding but incorrect or fabricated information by a generative model.}
{An LLM might invent a nonexistent academic reference in its response.}
{Hallucinations undermine trust and can have legal, safety, or reputational consequences if unmitigated.}

\glossentry{Human-in-the-loop (HITL)}
{A system design where humans review, correct, or override AI decisions.}
{A content moderation system may flag posts for human review instead of automatically removing them.}
{HITL improves safety, reduces errors, and ensures compliance in sensitive applications.}

\glossentry{KPI (Key Performance Indicator)}
{A measurable metric that indicates how well a system or process is achieving its objectives.}
{A fraud detection system’s KPI might be the percentage of fraudulent transactions caught without false positives.}
{KPIs connect technical performance to business outcomes, enabling meaningful evaluation of AI’s value.}

\glossentry{Lineage}
{The traceable record of how data, models, and artifacts were created, transformed, and used over time.}
{Lineage tracking can show which dataset version and preprocessing pipeline produced a given model version.}
{Lineage supports reproducibility, auditing, and regulatory compliance.}

\glossentry{LLMOps (Large Language Model Operations)}
{A specialization of MLOps focused on developing, deploying, and maintaining large language models and generative AI systems.}
{LLMOps workflows might manage prompt templates, fine-tuned checkpoints, and hallucination monitoring.}
{LLMOps addresses the unique challenges of generative AI, from prompt versioning to content safety.}

\glossentry{MLOps (Machine Learning Operations)}
{The set of practices, tools, and cultural approaches that unify machine learning system development and operations.}
{A team using MLOps might automate data ingestion, model training, deployment, and monitoring in a continuous workflow.}
{MLOps is foundational for scaling AI from prototypes to reliable, maintainable production systems.}

\glossentry{Model Card}
{A standardized documentation format that describes a model’s purpose, data, performance, limitations, and ethical considerations.}
{A vision model’s model card might include training data sources, known biases, accuracy by demographic, and appropriate use cases.}
{Model cards improve transparency, governance, and stakeholder communication.}

\glossentry{Observability}
{The ability to understand a system’s internal state based on external outputs such as logs, metrics, and traces.}
{Observability tools might track inference latency, model confidence scores, and data drift metrics over time.}
{Observability enables proactive monitoring, rapid debugging, and informed decision-making in production AI systems.}

\glossentry{Pipeline}
{A sequence of automated steps for data processing, training, validation, deployment, and monitoring.}
{A training pipeline might ingest raw data, preprocess it, train a model, evaluate performance, and register the model.}
{Pipelines enforce reproducibility, scalability, and automation across the AI lifecycle.}

\glossentry{POC (Proof of Concept)}
{A small-scale experiment or prototype used to test feasibility before committing significant resources.}
{A team might build a POC chatbot using open-source models before investing in a production deployment.}
{POCs reduce risk and inform decision-making in early project stages.}

\glossentry{Post-deployment Monitoring}
{The continuous observation of model behavior, inputs, outputs, and outcomes after deployment.}
{Monitoring might detect performance degradation due to data drift or unexpected user behavior.}
{Post-deployment monitoring is essential for sustaining performance, safety, and compliance over time.}

\glossentry{Prompt Engineering}
{The practice of crafting and refining prompts to guide the behavior of large language models (LLMs).}
{A customer support chatbot might respond more accurately when prompted with structured instructions and examples.}
{Effective prompt engineering is essential for controlling generative AI systems and aligning them with desired outcomes.}

\glossentry{RAG (Retrieval-Augmented Generation)}
{A technique that combines a generative model with an external knowledge retrieval step to improve accuracy and grounding.}
{A legal assistant LLM may retrieve relevant case law from a database before generating an answer.}
{RAG improves factual reliability and reduces hallucinations in generative AI applications.}

\glossentry{Retraining Cycle}
{The periodic process of retraining models to incorporate new data or address drift.}
{A recommendation system might retrain weekly using the latest user interaction data.}
{Regular retraining ensures models remain accurate, relevant, and aligned with changing realities.}

\glossentry{Rollback}
{Reverting a deployed model or service to a previous version after e.g. detecting performance or compliance issues.}
{A rollback might occur if a new model version causes error rates to spike in production.}
{Rollbacks are a safety mechanism that protect business continuity and reliability.}

\glossentry{Shadow Deployment}
{A deployment strategy where a new model runs in parallel with the current production model but does not affect outputs.}
{A bank might run a new fraud detection model in shadow mode to compare its decisions against the production model without impacting customers.}
{Shadow deployments allow safe real-world evaluation of new models before switching them into production.}

\glossentry{SLA (Service Level Agreement)}
{A contractual or internal agreement specifying measurable performance targets for a service, such as availability, latency, or throughput.}
{An inference API might have an SLA guaranteeing 99.9\% uptime and response times under 200 ms.}
{SLAs define expectations and accountability between teams or organizations, ensuring that deployed AI services meet operational requirements and business needs.}

\glossentry{SLO (Service Level Objective)}
{A specific, measurable performance target used internally to track service quality and adherence to SLAs.}
{A model-serving service may define an SLO of 99.95\% uptime, which supports achieving the 99.9\% SLA promised to customers.}
{SLOs provide engineering teams with concrete, actionable goals that align system performance with business commitments.}

\glossentry{Synthetic Data}
{Artificially generated data used to train or test models without exposing real sensitive data.}
{A healthcare company might use synthetic patient records to test models while preserving privacy.}
{Synthetic data expands training options, mitigates privacy risks, and supports regulatory compliance.}

\glossentry{Value Drift}
{The divergence between an AI system’s outputs and the evolving sources of business or societal value it was meant to serve.}
{A recommendation system optimized for engagement may harm user trust over time, reducing actual value.}
{Monitoring for value drift ensures AI remains aligned with strategic goals as conditions change.}

\glossentry{Value Hypothesis}
{An explicit assumption about how an AI system will generate value for the organization or users.}
{A predictive maintenance model is expected to reduce unplanned downtime by 30\%.}
{A clear value hypothesis guides design decisions and provides a basis for measuring success.}

\bibliographystyle{IEEEtran}
\bibliography{bibliography}

\end{document}